\documentclass[smallextended]{svjour3}

\usepackage{amssymb,amsmath}
\usepackage{subfig} 
\usepackage{array,booktabs}

\usepackage{graphicx}
\usepackage{wrapfig}

\usepackage{tikz} 
\usetikzlibrary{arrows,automata,shapes} 
\tikzstyle{every picture}+=[remember picture]
\everymath{\displaystyle}

\usepackage[ruled]{algorithm}
\usepackage{algpseudocode}

\usepackage{url}
\urldef{\mailsts}\path|{dreossi, sseshia}@berkeley.edu|    
\urldef{\mailsa}\path|alex.r.donze@gmail.com|    

\newcommand{\comment}[1]{}
\newcommand{\revised}[1]{#1}

\newcommand{\reals}{\mathbb{R}}
\newcommand{\bools}{\mathbb{B}}
\newcommand{\naturals}{\mathbb{N}}
\newcommand{\rationals}{\mathbb{Q}}

\newcommand{\stsp}{S}
\newcommand{\insp}{U}
\newcommand{\tsp}{T}
\newcommand{\simu}{sim}

\newcommand{\va}{\mathbf{a}}

\newcommand{\vu}{\mathbf{u}}

\newcommand{\vx}{\mathbf{s}} 

\newcommand{\pred}{\sigma}
\newcommand{\U}[1]{U_{#1}}

\newcommand{\F}[1]{F_{#1}}
\newcommand{\G}[1]{G_{#1}}

\newcommand{\rob}[3]{\rho(#1,#2,#3)}

\newcommand{\satsig}[3]{\mathcal{X}(#1,#2,#3)}

\newcommand\norm[1]{\left\lVert#1\right\rVert}

\newcommand{\featsp}{X}
\newcommand{\labsp}{Y}
\newcommand{\abssp}{A}
\newcommand{\class}{f}
\newcommand{\absclass}{\tilde{\class}}
\newcommand{\cfeatsp}{\tilde{\featsp}}
\newcommand{\cabssp}{\tilde{\abssp}}

\newcommand{\err}[2]{err_{#1}(#2)}

\newcommand{\fp}[2]{fp_{#1}(#2)}
\newcommand{\fn}[2]{fn_{#1}(#2)}

\newcommand{\feat}{\mathbf{x}}
\newcommand{\lab}{y}
\newcommand{\afeat}{\va}
\newcommand{\ifeat}[1]{\feat^{(#1)}}
\newcommand{\ilab}[1]{\lab^{(#1)}}
\newcommand{\iafeat}[1]{\afeat^{(#1)}}

\newcommand{\absf}{\alpha}
\newcommand{\conf}{\gamma}

\newcommand{\disc}[1]{D(#1)}

\newcounter{myctr}

\newenvironment{myitemize}{\begin{list}{$\bullet$}
{\setlength{\topsep}{1mm}\setlength{\itemsep}{0.25mm}
\setlength{\parsep}{0.1mm}
\setlength{\itemindent}{0mm}\setlength{\partopsep}{0mm}
\setlength{\labelwidth}{15mm}
\setlength{\leftmargin}{4mm}}}{\end{list}}

\begin{document}


\title{Compositional Falsification of Cyber-Physical Systems with Machine Learning Components\thanks{This work is funded in part by the DARPA BRASS program under agreement number FA8750-16-C-0043, NSF grants CNS-1646208, CNS-1545126, CCF-1837132, and CCF-1139138, the DARPA Assured Autonomy program, 
Toyota under the iCyPhy center, Berkeley Deep Drive, and by TerraSwarm, one of six centers of STARnet, a Semiconductor Research Corporation program sponsored by MARCO and DARPA. All the contributions of the second author with the exception of those to Sec. 5.2 occurred while he was affiliated with UC Berkeley. We gratefully acknowledge the support of NVIDIA Corporation with the donation of the Titan Xp GPU used for this research.}}

\titlerunning{Compositional Falsification of CPSML}

%
%
\author{Tommaso Dreossi \and Alexandre Donz\'e \and Sanjit A. Seshia}
\authorrunning{Dreossi, Donz\'e, Seshia}

\institute{T. Dreossi \at
	University of California, Berkeley\\
	\email{dreossi@berkeley.edu}
	\and
	A. Donz\'e \at
	Decyphir, Inc.\\
	\email{alex.r.donze@gmail.com}
	\and
	S. A. Seshia \at
	University of California, Berkeley\\
	\email{sseshia@berkeley.edu}
}

%
%

\maketitle

\begin{abstract}
Cyber-physical systems (CPS), such as automotive systems, are starting to include
sophisticated machine learning (ML) components. Their correctness, therefore, depends on
properties of the inner ML 
modules. While learning algorithms aim to generalize from examples, they are only 
as good as the examples provided, and 
recent efforts have shown that they can produce inconsistent output under small
adversarial perturbations. This raises the question: can the output from learning components
lead to a failure of the entire CPS? In this work, we address this question by
formulating it as a problem of
falsifying signal temporal logic specifications for CPS with ML components. 
We propose a compositional falsification framework where a temporal
logic falsifier and a machine learning analyzer cooperate with the aim  
of finding falsifying executions of the considered model. The efficacy
of the proposed technique is shown on an automatic emergency braking system
model with a perception component based on deep neural networks.
\end{abstract}

\section{Introduction}\label{sec:introduction}

Over the last decade, machine learning (ML) algorithms have achieved impressive results providing solutions to practical large-scale
problems (see, e.g.,~\cite{blum1997selection,michalski2013machine,jia2014caffe,hinton2012deep}).
Not surprisingly, ML is being used in 
cyber-physical systems (CPS) --- systems that are integrations of
computation with physical processes. 
For example, semi-autonomous vehicles employ Adaptive Cruise Controllers (ACC)
or Lane Keeping Assist Systems (LKAS) that rely heavily on image classifiers
providing input to the software controlling electric and mechanical 
subsystems (see, e.g.,~\cite{nvidiaself-arxiv16}).
The safety-critical nature of such systems involving ML
raises the need for formal methods~\cite{SeshiaS16}.
In particular, how do we systematically find bugs in such systems?

We formulate this question as the falsification problem for CPS models with 
ML components (CPSML): given a formal specification $\varphi$ 
(say in a formalism such as signal temporal logic~\cite{maler2004monitoring}) 
and a CPSML model $M$, find an input for which $M$ does {\it not} satisfy $\varphi$. A falsifying 
input generates a counterexample trace that reveals a bug.
To solve this problem, multiple challenges must be tackled. 
First, the input space to be searched can be intractable.
For instance, a simple model of a semi-autonomous car already involves
several control signals (e.g., the angle of the acceleration pedal, steering angle) 
and other rich sensor input (e.g., images captured by a camera, LiDAR, RADAR). 
Second, the formal verification of ML components is a difficult,
and somewhat ill-posed problem due to
the complexity of the underlying ML algorithms,
large feature spaces, and the lack of consensus on a formal definition
of correctness of an ML component. The last point is an especially
tricky challenge for ML-based perception; see~\cite{SeshiaS16,seshia-atva18} for a longer discussion on specification of ML components. 
Third, CPSML are often designed using languages 
such as C, C++, or Simulink for which clear semantics are not given,
and involve third-party components that are opaque or poorly-specified. 
This obstructs the development of formal methods for the analysis 
of CPSML models and may force one 
to treat them as gray-boxes or black-boxes.
Hence, we need a technique 
to systematically analyze ML components within the context of a CPS
that can handle all three of these challenges.

In this paper, we propose a new framework for the falsification of CPSML
addressing the issues described above. 
Our technique is {\em compositional (modular)} in that it divides the 
search space for falsification
into that of the ML component and of the
remainder of the system, while 
establishing a connection between the two.
The obtained projected search spaces are respectively analyzed by 
a temporal logic falsifier (``CPS Analyzer'') and a machine learning
analyzer (``ML analyzer'') that cooperate to search for
a behavior of the closed-loop system that violates the property $\varphi$.
This cooperation mainly comprises a sequence of input space projections,
passing information about interesting regions in the input space
of the full CPSML model to identify a sub-space of the input space
of the ML component. The resulting projected input space of the ML
component is typically smaller than the full input space. Moreover,
misclassifications in this space can be mapped back to smaller subsets
of the CPSML input space in which 
counterexamples are easier to find.
Importantly, our approach can handle {\em any machine learning technique},
including the methods based on deep neural networks~\cite{hinton2012deep} 
that have proved effective in many recent applications.
The proposed ML Analyzer is a tool that analyzes the
input space for the ML classifier and determines
a region of the input space that could be relevant for the full 
cyber-physical system's correctness. More concretely,
the analyzer identifies sets of misclassifying
features, i.e., inputs that ``fool'' the ML algorithm. The analysis 
is performed by considering subsets of parameterized features spaces
that are used to approximate the ML components by simpler functions. 
The information gathered by the temporal logic falsifier and the
ML analyzer together reduce the search space,
providing an efficient approach to falsification for CPSML models.

\begin{figure}
  	\begin{center}
		\includegraphics[width=0.8\textwidth]{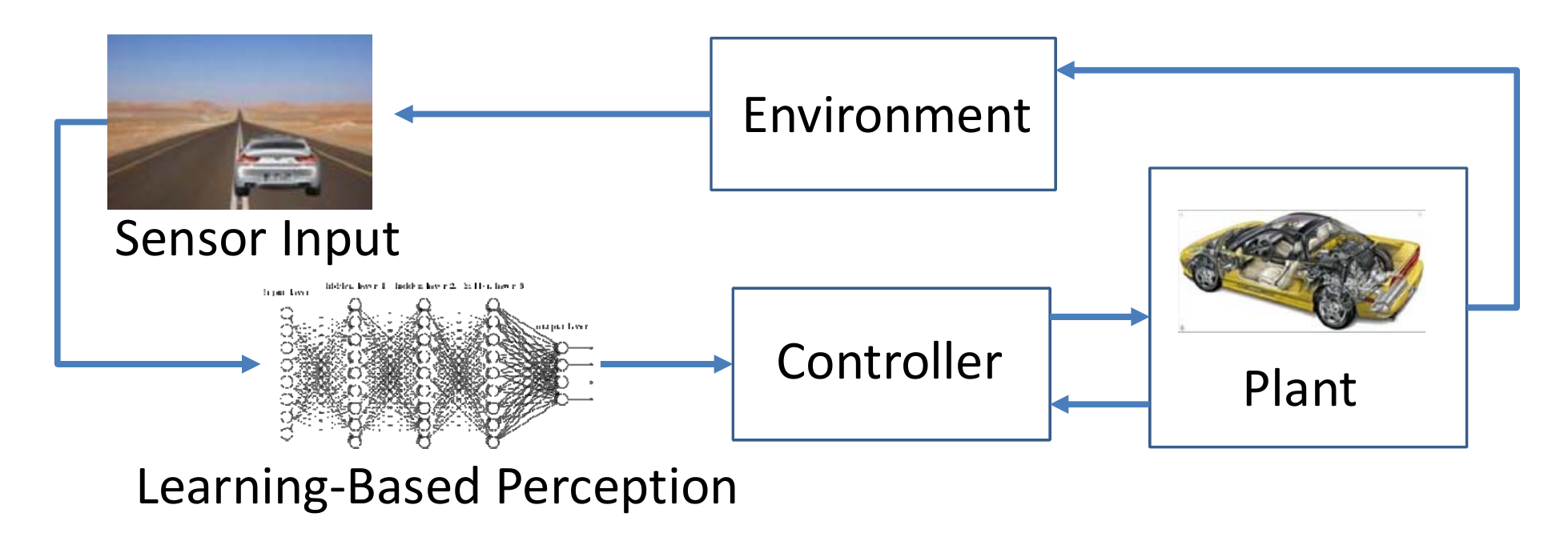}
	 \end{center}
	\caption{Automatic Emergency Braking System (AEBS) in closed loop. A machine learning based image classifier is used to perceive objects in the ego vehicle's frame of view.\label{fig:aebs_ex}}
\end{figure}

\begin{example}\label{ex:aebs}

As an illustrative example, let us consider a simple model of an Automatic Emergency Braking System (AEBS),
that attempts to detect objects in front of a vehicle and actuate
the brakes when needed to avert a collision.
Figure~\ref{fig:aebs_ex} shows the AEBS as a system
composed of a controller (automatic braking), a plant (vehicle sub-system under control, including transmission), and an advanced sensor (camera along
with an obstacle detector based on deep learning). The AEBS, when combined
with the vehicle's environment, forms a closed loop control system.
The controller regulates the acceleration and braking of the plant using the velocity of
the subject (ego) vehicle and the distance between it and an obstacle.
The sensor used to detect the obstacle includes a camera along with an image classifier based on deep neural networks.
In general, this sensor can provide noisy measurements due to incorrect image 
classifications which in turn can affect the correctness of the overall system.

Suppose we want to verify 
whether the distance between the subject vehicle and a preceding obstacle is always larger than
$5$ meters. Such a verification requires the exploration of a very 
large input space comprising 
the control inputs (e.g., acceleration and braking pedal angles) and the ML component's feature space
(e.g., all the possible pictures observable by the camera). 
The latter space is particularly large --- for example,
note that the feature space of RGB images of dimension 
$1000\times600$px (for an image classifier) 
contains $256^{1000\times 600 \times 3}$ elements.
\end{example}

At first, the input space of the model described in Example~\ref{ex:aebs} appears intractably large.
However, the twin notions of {\em abstraction} and {\em compositionality},
central to much of the advances in formal verification, can help address
this challenge. As mentioned earlier, we decompose the overall CPSML
model input space into two parts: (i) the input space of the ML component,
and (ii) the input space for the rest of the system -- i.e., the CPSML
model with an abstraction of the ML component.
A {\em CPS Analyzer} operates on the latter ``pure CPS'' input space,
while an {\em ML Analyzer} handles the former. 
The two analyzers communicate information as follows:
\begin{enumerate}
	\item The CPS Analyzer initially performs conservative analyses assuming abstractions of the ML component. In particular, consider two extreme abstractions --- a \lq\lq perfect ML classifier\rq\rq\ (i.e., all feature vectors are correctly classified), and a ``completely-wrong ML classifier'' (all feature vectors are misclassified). Abstraction permits the CPS Analyzer to operate on a lower-dimensional input space (the ``pure CPS'' one) and identify a region in this space that may be
affected by the malfunctioning of some ML modules -- a so-called ``region of interest'' or ``region of uncertainty.''
This region is communicated to the ML Analyzer.

	\item The ML Analyzer projects the region of uncertainty (ROU) onto its input space, and performs a detailed analysis of that input sub-space. Since this detailed analysis uses only the ML classifier (not the full CPSML model), it is a more tractable problem. In this paper, we present a novel
sampling-based approach to explore the input sub-space for the ML
component. We can also leverage other advances in analysis of machine 
learning systems operating on rich sensor inputs and for applications
such as autonomous driving (see the related work section that follows).

	\item When the ML Analyzer finds interesting test cases (e.g., those that trigger misclassifications of inputs whose labels are easily inferred), it communicates that information back to the CPS Analyzer, which checks whether the ML misclassification can lead to a system-level safety violation (e.g., a collision). If yes, we have found a system-level counterexample.
If not, the ROU is updated and the revised ROU passed back to the ML Analyzer. 

\end{enumerate}
The communication between the CPS Analyzer and ML Analyzer continues
until either we find a system-level counterexample, or we run out of
resources. For the class of CPSML models we consider, including those
with highly non-linear dynamics and even black-box components, one
cannot expect to prove system correctness.
We focus on specifications in Signal Temporal Logic (STL), and
for this reason use a temporal logic falsifier, Breach~\cite{donze2010breach}, 
as our CPS Analyzer; however, other specification formalisms and tools
may also be used and our framework is largely agnostic to these choices.
Even though temporal logic falsification is a mature technology with
initial industrial adoption (e.g.,~\cite{yamaguchi-fmcad16}),
several technical challenges remain.
First, we need to construct the validity domain of an STL
specification --- the input sub-space where the property is 
satisfied --- for a CPSML model with abstracted (correct/incorrect) 
ML components, and identify the region of uncertainty (ROU).
Second, we need a method to relate the ROU 
to the feature space of the ML modules. 
Third, we need to systematically analyze the feature space of 
the ML component
with the goal of finding feature vectors leading to misclassifications.
We describe in detail in Sections~\ref{sec:framework} 
and~\ref{sec:MLanalysis} how we tackle these challenges.

\comment{
However, we can observe some interesting aspects of the relationship between the ``pure CPS'' input space
and its ML feature space:
\begin{enumerate}
	\item Under the assumption of ``perfect ML components'' (i.e., all feature vectors are correctly classified),
		we can study the CPSML model on a 
		lower-dimensional input space (the ``pure CPS'' one) and identify regions of values that satisfy the specification but might be
		affected by the malfunctioning of some ML modules;
	\item Instead of verifying the ML components on their whole feature spaces, we can focus only on those
		features related to the non-robust input values identified in the previous step, and
	\item If we are able to determine misclassifications on the restricted feature space,
		then we can relate them back to CPSML input space, thus focusing the falsification on a smaller
		input space.
\end{enumerate}
} 

In summary, the main contributions of this paper are:
\begin{myitemize}

\item A compositional framework for
the falsification of temporal logic properties of arbitrary 
CPSML models that works for any kind of machine learning classifier.
To our knowledge, we present the first approach to falsifying temporal
logic properties of closed-loop CPS with ML components, including
deep neural networks.

\item The first approach for formal analysis of systems that 
use ML for perception. In particular, we show how to use
a novel combination of {\em abstraction} and
{\em compositional reasoning} to scale falsification to the
higher-dimensional input spaces occurring for ML-based perception.
Our compositional framework is the first modular approach to
verification of CPSML, and the first to systematically 
deal with the specification challenge for ML-based perception
as described in~\cite{SeshiaS16}.

\item A machine learning analyzer that identifies
misclassifications leading to system-level property violations, 
based on the following main ideas:
\begin{myitemize}
	\item[-] An input space parameterization used to abstract the feature space of the ML component and relate it to the CPSML input space;
	\item[-] Systematic sampling methods to explore the feature space of the ML component, and 
	\item[-] A classifier approximation method used to abstract
the ML component and identify misclassifications that can 
lead to executions of the CPSML that violate the temporal logic specification
(i.e., system-level counterexamples).
\end{myitemize}

\item
An experimental demonstration of the effectiveness of our approach 
on two instantiations of an Automatic Emergency Braking System (AEBS)
example with multiple deep neural networks trained for object
detection and classification, including some developed by experts
in the machine learning and computer vision communities.
\end{myitemize}
In Sec.~\ref{sec:experiments}, we give detailed experimental results
on an Automatic Emergency Braking System (AEBS)
involving an image classifier for obstacle detection
based on deep neural networks developed and trained
using leading software packages --- 
AlexNet developed with Caffe~\cite{jia2014caffe} 
and Inception-v3 developed with TensorFlow~\cite{tensorflow2015}. 
In this journal version of our original conference paper~\cite{dreossi-nfm17},
we also present a new case study, an AEBS deployed within 
the Udacity self-driving car simulator~\cite{udacitySim-www} trained
on images generated from the simulator.

\subsection*{Related Work}

The verification of both CPS and ML algorithms have attracted several research efforts,
and we focus here on the most closely related work.
Techniques for the falsification of temporal logic specifications against CPS models
have been implemented based on nonlinear optimization methods and stochastic search strategies
(e.g., Breach~\cite{donze2010breach}, S-TaLiRo~\cite{annpureddyLFS11}, 
RRT-REX~\cite{dreossi2015efficient}, C2E2~\cite{duggirala2015c2e2}).
While the verification of ML programs is less well-defined~\cite{SeshiaS16},
recent efforts~\cite{szegedy2013intriguing} show how
even well trained neural networks can be sensitive to small adversarial perturbations, i.e., small intentional modifications that
lead the network to misclassify the altered input with large confidence.
Other efforts have tried to characterize the correctness of neural networks 
in terms of risk~\cite{vapnik1991principles}
(i.e., probability of misclassifying a given input) or 
robustness~\cite{fawzi2015analysis,carlini-ieeesp17} 
(i.e., a minimal perturbation leading to a misclassification),
while others proposed methods to generate pictures~\cite{nguyen2015deep,dreossi-rmlw17} 
or perturbations~\cite{moosavi2015deepfool,nguyen2015deep}
including methods based on satisfiability modulo theories (SMT) 
(e.g.,~\cite{huang-arxiv16}) 
in such a way to ``fool'' neural networks. 
These methods, while very promising, are limited
to analyzing partial specifications of the ML components in isolation, 
and not in the context of a complex, closed-loop
cyber-physical system. 
To the best of our knowledge, 
our work is the first to address the verification of end-to-end temporal logic
properties of CPSML---the combination
of CPS and ML systems.
The work that is closest in spirit to ours is that on DeepXplore~\cite{pei-sosp17},
where the authors present a whitebox software testing approach
for deep learning systems. However, there are some important
differences: their work performs a detailed analysis of the learning software, whereas
ours analyzes the entire closed-loop CPS while delegating the software
analysis to the machine learning analyzer. Further, we consider temporal
logic falsification whereas their work uses software and neural
network coverage metrics. It may be interesting to see how these
approaches can be combined.
We believe our approach is the first step towards performing a
more ``semantic'' adversarial analysis of systems that employ machine learning,
including deep learning~\cite{dreossi-cav18}.

\section{Background}
\label{sec:background}

\subsection{CPSML Models}
\label{sec:cps}

In this work, we consider models of cyber-physical systems with machine learning components (CPSML).
We assume that a system model is given as a gray-box simulator defined as a tuple $M= (\stsp, \insp, \simu)$,
where $\stsp$ is a set of system states, $\insp$ is a set of input values, and
$\simu: \stsp \times \insp \times \tsp  \to \stsp$ is a simulator that maps 
a state $\vx(t_k) \in \stsp$ and input value $\vu(t_k) \in \insp$ at time $t_k \in \tsp$ to a new 
state $\vx(t_{k+1}) = \simu(\vx(t_k), \vu(t_k),t_k)$, where $t_{k+1} = t_k + \Delta_k$ for a 
time-step $\Delta_k \in \rationals_{> 0}$.

Given an initial time $t_0 \in \tsp$, an initial state $\vx(t_0) \in \stsp$, a sequence of time-steps
$\Delta_0, \dots, \Delta_n \in \rationals_{> 0}$, and a sequence of input values $\vu(t_0), \dots, \vu(t_n) \in \insp$,
a simulation trace of the model $M= (\stsp, \insp, \simu)$ is a sequence:
$$(t_0,\vx(t_0),\vu(t_0)),(t_1,\vx(t_1),\vu(t_1)), \dots, (t_n,\vx(t_n),\vu(t_n))$$
where $\vx(t_{k+1}) = \simu(\vx(t_k),\vu(t_k),\Delta_k)$ and $t_{k+1} = t_{k} + \Delta_k$ for $k = 0,\dots, n$.

The gray-box aspect of the CPSML model is that we assume some knowledge
of the internal ML components. Specifically, these components,
termed \emph{classifiers},
are functions $\class : \featsp \to \labsp$ that assign to their input 
\emph{feature vector} $\feat \in \featsp$ a \emph{label} $\lab \in \labsp$,
where $\featsp$ and $\labsp$ are a feature and label space,
respectively. Without loss of generality, we focus on binary classifiers whose label space is $\labsp = \{0,1\}$.
An ML algorithm selects 
a classifier using a training set $\{(\ifeat{1},\ilab{1}),\dots,(\ifeat{m},\ilab{m})\}$
where the $(\ifeat{i},\ilab{i})$ are labeled examples with $\ifeat{i} \in \featsp$ and
$\ilab{i} \in \labsp$, for $i = 1,\dots, m$.
The quality of a classifier can be estimated on a test set of examples 
comparing the classifier predictions 
against the labels of the examples. Precisely, for a given 
test set $T = \{(\ifeat{1},\ilab{1}),\dots,(\ifeat{l},\ilab{l})\}$,
the number of false positives $\fp{f}{T}$ and false negatives $\fn{f}{T}$ of 
a classifier $f$ on $T$ are defined as:
\begin{equation}
	\begin{split}
		\fp{f}{T} = &\ \mid \{ \ifeat{i} \in T \mid \class(\ifeat{i}) = 1 \text{ and } \ilab{i} = 0 \} \mid \\
		\fn{f}{T} = &\ \mid \{ \ifeat{i} \in T \mid \class(\ifeat{i}) = 0 \text{ and } \ilab{i} = 1 \} \mid \\
	\end{split}
\end{equation}
The error rate of $f$ on $T$ is given by:
\begin{equation}
	\err{f}{T} = (\fp{f}{T}+\fn{f}{T}) / l 
\end{equation}
A low error rate implies good predictions of the classifier $f$ on the test 
set $T$.

\subsection{Signal Temporal Logic}

We consider Signal Temporal Logic~\cite{maler2004monitoring}
(STL) as the language to specify properties to be verified against a CPSML model.
STL is an extension of linear temporal logic (LTL) suitable for the 
specification of properties of CPS.

A \emph{signal} is a function $s : D \to S$, with $D \subseteq \reals_{\geq 0}$ 
an interval and either $S \subseteq \bools$ or $S \subseteq \reals$,
where $\bools = \{\top, \bot\}$ and $\reals$ is the set of reals. 
Signals defined on $\bools$ are called \emph{booleans}, while
those on $\reals$ are said \emph{real-valued}. 
A \emph{trace} $w = \{s_1,\dots, s_n\}$ is a finite set of real-valued signals
defined over the same interval $D$.

Let $\Sigma = \{\pred_1, \dots, \pred_k \}$ be a finite set of predicates $\pred_i : \reals^n \to \bools$,
with $\pred_i \equiv p_i(x_1, \dots, x_n) \lhd 0$, $\lhd \in \{ <, \leq \}$, and $p_i : \reals^n \to \reals$
a function in the variables $x_1, \dots, x_n$.

An STL formula is defined by the following grammar:
\begin{equation}
	\varphi := \pred \, | \, \neg \varphi \, | \, 
                    \varphi \wedge \varphi \, | \, \varphi \U{I} \varphi
\end{equation}
where $\pred \in \Sigma$ is a predicate and $I \subset \reals_{\geq 0}$ is a closed
non-singular interval. Other common temporal operators
can be defined as syntactic abbreviations in the usual way, like for instance
$\varphi_1 \vee \varphi_2 := \neg ( \neg \varphi_1 \wedge \varphi_2 )$,
$\F{I} \varphi := \top \U{I} \varphi$, or $\G{I}\varphi := \neg \F{I} \neg \varphi$.
Given a $t \in \reals_{\geq 0}$, a shifted interval $I$
is defined as $t + I = \{t + t' \mid t' \in I\}$.


\begin{definition}[Qualitative semantics]
	Let $w$ be a trace, $t \in \reals_{\geq 0}$, and $\varphi$ be an STL formula.
	The \emph{qualitative semantics} of $\varphi$ is inductively defined as follows:
	\begin{equation}
		\begin{split}
			w,t \models  \pred \text{ iff } 			&\pred(w(t)) \text{ is true} \\
			w,t \models \neg\varphi \text{ iff } 		& w,t \not\models \varphi \\
			w,t \models \varphi_1 \wedge \varphi_2 \text{ iff }	& w,t \models \varphi_1 \text{ and } w,t \models \varphi_2 \\
			w,t \models \varphi_1 \U{I} \varphi_2 \text{ iff }	& \exists t' \in t + I \text{ s.t. } w,t' \models \varphi_2 \text{ and } \forall t'' \in [t,t'], w,t'' \models \varphi_1 \\
		\end{split}
	\end{equation}
\end{definition}

A trace $w$ satisfies a formula $\varphi$ if and only if $w,0 \models \varphi$, in short $w \models \varphi$.
For given signal $w$, time instant $t\in\reals_{\geq 0}$, and STL formula $\varphi$, the \emph{satisfaction
signal} $\satsig{w}{t}{\varphi}$ is $\top$ if $w,t \models \varphi$,  $\bot$ otherwise.

Given a CPSML model $M = (\stsp, \insp, \simu)$, $M \models \varphi$ if
every simulation trace of $M$ satisfies $\varphi$.

\begin{definition}[Quantitative semantics]
	Let $w$ be a trace, $t \in \reals_{\geq 0}$, and $\varphi$ be an STL formula.
	The \emph{quantitative semantics} of $\varphi$ is defined as follows:
	\begin{equation}
		\begin{split}
			\rob{p(x_1, \dots, x_n) \lhd 0}{w}{t} = &\ p(w(t)) \text{ with } \lhd \in \{<,\leq \} \\
			\rob{\neg\varphi}{w}{t} = &\ -\rob{\varphi}{w}{t} \\
			\rob{\varphi_1 \wedge \varphi_2}{w}{t} = &\ \min( \rob{\varphi_1}{w}{t}, \rob{\varphi_2}{w}{t} ) \\
			\rob{\varphi_1 \U{I} \varphi_2}{w}{t} = &\ \sup_{t' \in t+I} \min( \rob{\varphi_2}{w}{t'}, \inf_{t''[t,t']} \rob{\varphi_1}{w}{t''} ) \\
		\end{split}
	\end{equation}
\end{definition}
The \emph{robustness} of a formula $\varphi$ with respect to a trace $w$ is the signal $\rob{\varphi}{w}{\cdot}$.

{\revised{Quantitative semantics helps to determine how robustly a formula is satisfied.
Intuitively, the quantitative evaluation of a formula provides a real value representing
the distance to satisfaction or violation.
The  quantitative  and  qualitative  semantics  are  connected. Specifically, it holds that 
$\rob{\varphi}{w}{t} > 0$ if and only if $w,t \models \varphi$~\cite{donze2013efficient}.}}

Given a CPSML model $M = (\stsp, \insp, \simu)$, 
and a temporal logic formula $\varphi$,
the \emph{validity domain} of $\varphi$ for model $M$ is the subset of
$\insp$ for which traces of $M$ satisfy $\varphi$.
We denote the validity domain by $\insp_{\varphi}$; the remaining
set of inputs $\insp \setminus \insp_{\varphi}$ is denoted by
$\insp_{\neg\varphi}$.
{\revised{Note that there are no limitations on the dimensionality of a validity domain. It can potentially characterize
single initial conditions as well as entire input traces. Usually
when we speak of validity domains as sets of traces, we represent
them using a suitable finite parameterization of traces.}}
Simulation-based verification tools (such as~\cite{donze2010breach})
can approximately compute validity domains via sampling-based methods.

\section{Compositional Falsification Framework}
\label{sec:framework}

In this section, we formalize the falsification problem for STL specifications against CPSML models,
define our compositional falsification framework, and show its functionality on the AEBS system of
Example~\ref{ex:aebs}.



\begin{definition}[Falsification of CPSML]
Given a model $M = (\stsp,\insp,\simu)$ and an STL specification $\varphi$,
find an initial state $\vx(t_0) \in \stsp$ and a sequence of input values
$\vu = \vu(t_0), \dots, \vu(t_n) \in \insp$ such that the trace of states
$w = \vx(t_0),\dots, \vx(t_n)$
generated by the simulation of $M$ from $\vx(t_0) \in \stsp$
under $\vu$ does not satisfy $\varphi$,
i.e., $w \not\models \varphi$.
We refer to such $(\vx(t_0),\vu)$ as \emph{counterexamples} for $\varphi$.
The problem of finding a counterexample is often called
the \emph{falsification problem}.
\label{defn:fals_prob}
\end{definition}

We now present the compositional framework for the falsification of STL formulas against CPSML models.
Intuitively, the proposed method decomposes a given model into two parts: (i) an abstraction of the CPSML model
under the assumption of perfectly correct ML modules, and (ii) its actual ML components. The two parts 
are separately analyzed, the first by a temporal logic falsifier that builds the validity domain with respect to the given specification,
the second by an ML analyzer that identifies sets of feature vectors that are misclassified by the ML components.
Finally, the results of the two analyses are composed and projected back to a targeted input subspace of the original CPSML model
where counterexamples can be found by invoking a temporal logic falsifier.
We next formalize this procedure.

Let $M = (\stsp,\insp,\simu)$ be a CPSML model and $\varphi$ be an STL specification.
$U_\varphi$ is the validity domain of $\varphi$ for $M$. Thus, for falsification, we wish to
find an element of $U_{\neg \varphi}$. The challenge is that the dimensionality of
the input space $U$ is very high.

We address this challenge through a combination of {\em abstraction}
and {\em compositional reasoning}.
Consider creating an ``optimistic'' abstraction $M^+$ of $M$:
in other words, $M^+$ is a version of $M$ with perfect ML components, 
that is,
every feature vector of the ML feature space is correctly classified. Let us denote
by $ml$ the unabstracted ML components of the model $M$.

Under the assumption of correct ML components, the lower-dimensional input space
of $M^+$ can be analyzed by constructing the validity domain of $\varphi$, that is 
the partition of the input space of $M^+$ 
into the sets $U_\varphi^+$ and $U_{\neg\varphi}^+$
that do and do not satisfy $\varphi$, respectively.
The set $U_{\neg \varphi}^+$ comprises inputs on which the CPSML model violates
$\varphi$ even if the ML component operates perfectly, i.e., the system perceives
its environment perfectly.
For falsification, we are interested in identifying points in $U_{\varphi}^+$
that correspond to environments in which a misclassification produced by $ml$
can result in $M$ violating $\varphi$.
%
This corresponds to analyzing the behavior of the ML components $ml$ on 
inputs corresponding to the set $U_\varphi^+$.
We refer to this step as the ML analysis. It can be seen as a procedure
for finding a subset $U^{ml} \subseteq U$ of input values mapping to feature
vectors that are misclassified by the ML components $ml$.
It is important to note that the input space 
of the CPS model $M^+$ and the feature spaces of the ML modules $ml$ are 
different; thus, the ML analyzer must adapt and relate the two 
different spaces. This important step will be clarified in
Section~\ref{sec:MLanalysis}.

\revised{
Finally, the set $U^{ml}$ generated by the decomposed 
analysis of the CPS model and its ML components targets a small set of input values that 
are misclassified by the ML modules and are likely to falsify $\varphi$.
Thus, system-level counterexamples in $U^{ml} \subseteq U$ can be determined
by invoking a temporal logic falsifier on $\varphi$ against the original model $M$.
}

As explained below, we can pair the ``optimistic'' abstraction explained above 
with a ``pessimistic'' abstraction as well, so as to obtain
a further restriction of the input space.

\begin{algorithm}
	\caption{CPSML falsification scheme (one iteration between CPS Analyzer and ML Analyzer)}
  \label{algo:CPSMLfalsification}
  \begin{algorithmic}[1]
    	\Function{CompFalsfy}{$M, \varphi$}\Comment{$M$ CPSML, $\varphi$ STL specification}
	\State $[M^+,ml] \gets$\Call{Decompose}{$M$}\label{ln:decompose_pos}\Comment{$M^+$ -- perfect ML, $ml$ -- ML component}
	\State $[U_\varphi^+, U_{\neg\varphi}^+] \gets$\Call{ValidityDomain}{$M^+,U,\varphi$}\label{ln:valdom_pos}\Comment{Validity domain of $\varphi$ w.r.t. $M^+$}
	\State $[M^-,ml] \gets$\Call{Decompose}{$M$}\label{ln:decompose_neg}\Comment{$M^-$ -- wrong ML, $ml$ -- ML component}
	\State $[U_\varphi^-, U_{\neg\varphi}^-] \gets$\Call{ValidityDomain}{$M^-,U,\varphi$}\label{ln:valdom_neg}\Comment{Validity domain of $\varphi$ w.r.t. $M^-$}
	  \State $U_{rou} \gets U_\varphi^+ \setminus U_\varphi^-$  \label{ln:ROU} \Comment{Compute ROU}
	  \State $U^{ml} \gets$ \Call{MLAnalysis}{$ml,U_{rou}$}\label{ln:mlanalysis}\Comment{Find misclassified feature vectors in ROU}
\revised{	
	  \State $U^{ml}_{\neg\varphi} \gets $\Call{Falsify}{$M, U^{ml},\varphi$}\label{ln:fals}\Comment{Falsify property on original model with targeted inputs}
	  \State \Return $U_{\neg\varphi}^+ \cup U^{ml}_{\neg\varphi}$
}
      \EndFunction
  \end{algorithmic} 
\end{algorithm}

The compositional falsification procedure is formalized in Algorithm~\ref{algo:CPSMLfalsification} which shows one iteration of creating abstractions,
calling the CPS falsifier, and the ML analyzer. Figure~\ref{fig:compFalsify} shows the overall procedure which includes Algorithm~\ref{algo:CPSMLfalsification}.
\Call{CompFalsfy}{} receives as input a CPSML model $M$ and 
an STL specification $\varphi$, and returns a set of falsifying counterexamples.
At first, the algorithm decomposes $M$ into $M^+$ and $ml$, where $M^+$ is an abstract version of $M$ with ML components $ml$ that return perfect answers (classifications) (Line~\ref{ln:decompose_pos}).
Then, the validity domain of $\varphi$ with respect to the abstraction $M^+$ is computed 
by \Call{ValidityDomain}{} (Line~\ref{ln:valdom_pos}).
Next, the algorithm computes $M^-$ and $ml$ from $M$, where $M^-$ is an abstract version of $M$ with ML components $ml$ that always return wrong answers (misclassifications) (Line~\ref{ln:decompose_neg}).
Note that this step can be combined with Line~\ref{ln:decompose_pos},
but we leave it separate for clarity in the abstract algorithm specification.
Then, the validity domain of $\varphi$ with respect to the abstraction $M^-$ is computed 
by \Call{ValidityDomain}{} (Line~\ref{ln:valdom_neg}).
The region of uncertainty (ROU), where misclassifications of the
ML components can lead to violations of $\varphi$,
is then computed as $U_{rou}$ (Line~\ref{ln:ROU}).
From this, the subset of inputs to the CPSML model that are misclassified by 
$ml$ is identified by \Call{MLAnalysis}{} (Line~\ref{ln:mlanalysis}).
Finally, the targeted input set $U^{ml}$,
comprising the intersection of the sets identified by the decomposed 
analysis, is searched by a temporal logic falsifier on the original model $M$ (Line~\ref{ln:fals})
and the set of inputs that falsify the temporal logic formula
are returned.
It is important to note that the counterexamples we generate are for the
concrete CPSML model $M$, i.e., the one with the actual ML component,
not its optimistic or pessimistic abstraction.

\begin{figure}
  	\begin{center}
	\includegraphics[width=0.8\textwidth]{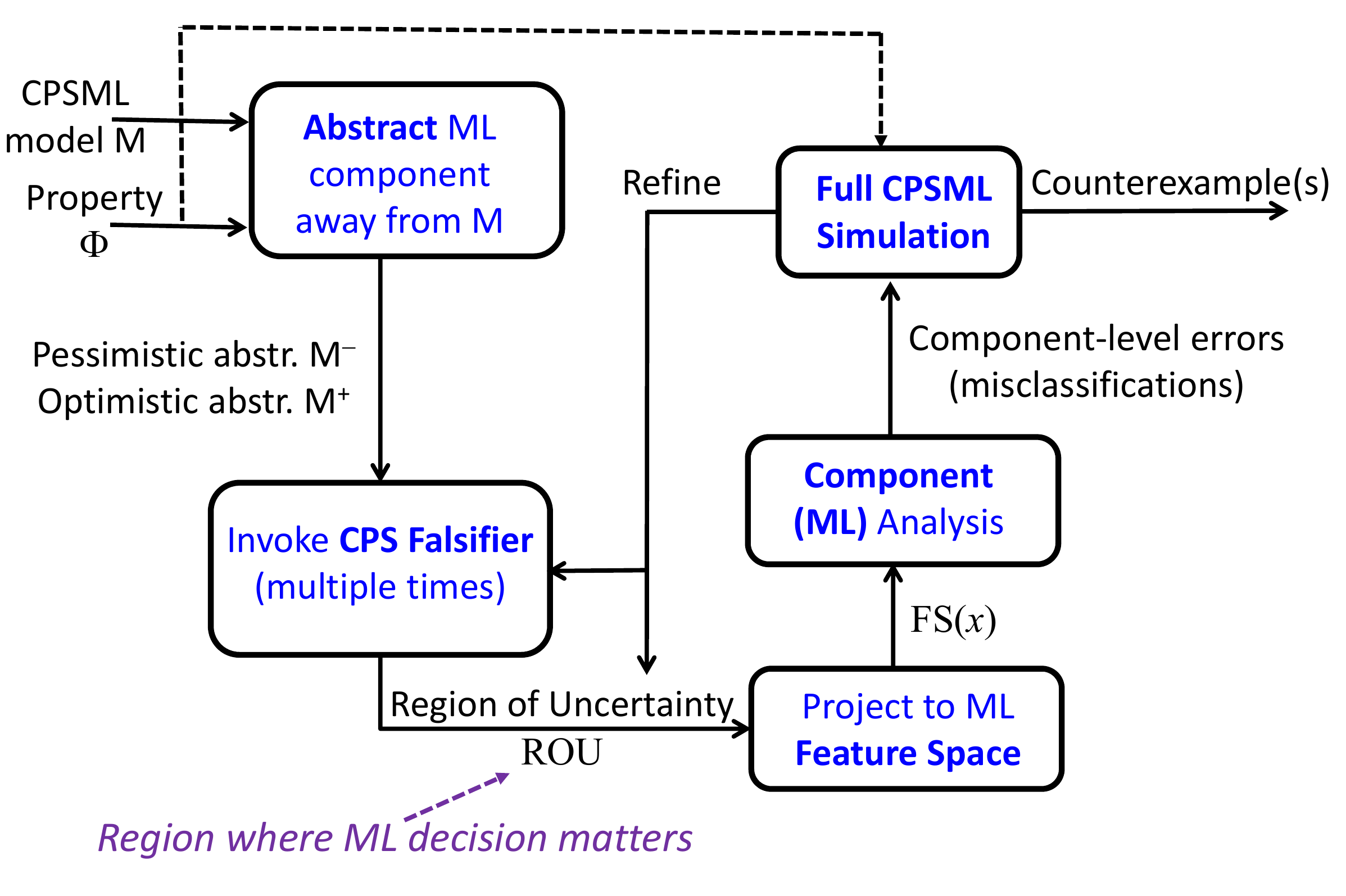}
	 \end{center}
	\caption{Compositional Falsification Technique. The method combines abstraction of the ML component with temporal logic falsification of the abstracted system and analysis of the ML component.\label{fig:compFalsify}}
\end{figure}

Note that the above approach can be implemented even without 
computing $M^-$ (Lines~\ref{ln:decompose_neg}-\ref{ln:ROU}), 
in which case the entire validity domain of $\varphi$ is considered
as the ROU. For simplicity, we will take this truncated approach
in the example described below. In Section~\ref{sec:experiments},
we will describe results on the AEBS case study with the full approach.

\comment{
The compositional falsification procedure is formalized in Algorithm~\ref{algo:CPSMLfalsification}.
\Call{CompFalsfy}{} receives as input a CPSML model $M$ and 
an STL specification $\varphi$, and returns a set of falsifying counterexamples.
At first, the algorithm decomposes $M$ into $M'$ and $ml$, where $M'$ is an abstract version of $M$ with perfectly
working ML modules, and $ml$ are the ML components of $M$ (Line~\ref{ln:decompose}).
Then, the validity domain of $\varphi$ with respect to the abstraction $M'$ is computed 
by \Call{ValidityDomain}{} (Line~\ref{ln:valdom}) and subsets of input that are misclassified by 
$ml$ are identified by \Call{MLAnalysis}{} (Line~\ref{ln:mlanalysis}).
Finally, the targeted input set $U_\varphi \cap U^{ml}$,
consisting in the intersection of the sets identified by the decomposed 
analysis, is searched by a temporal logic falsifier on the original model $M$ (Line~\ref{ln:fals})
and the set of inputs that falsify the temporal logic formula
are returned.
} 

\begin{figure}
  	\begin{center}
    		\includegraphics[width=0.5\textwidth]{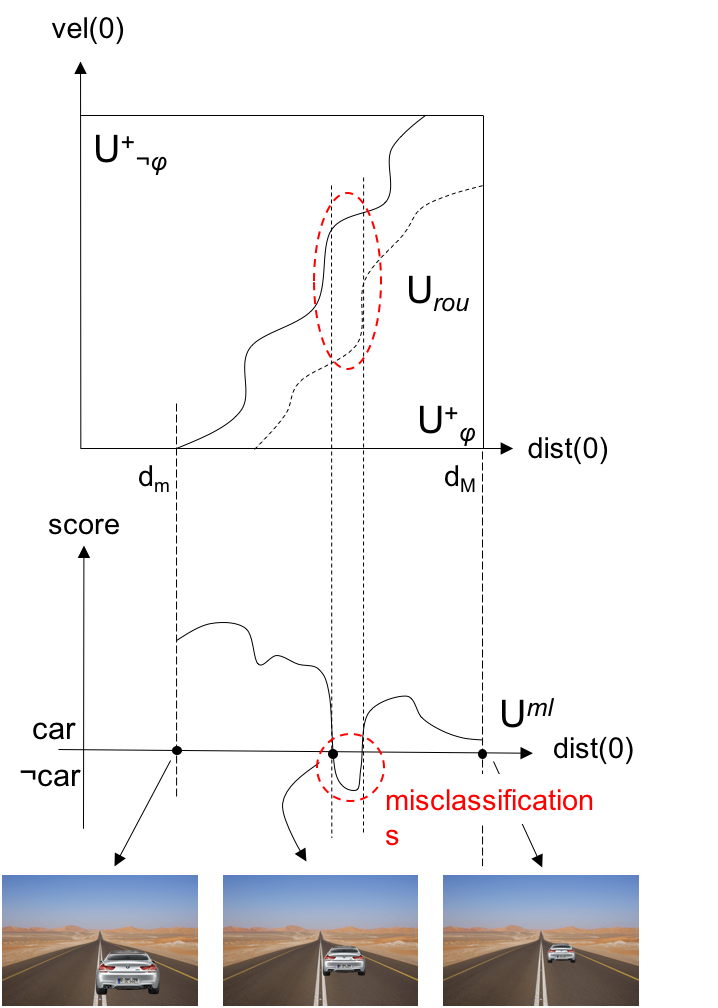}
	 \end{center}
  	\caption{Compositional falsification scheme on AEBS model. The ``score'' indicates the confidence level with which the classifier determines whether the image contains a car or not.\label{fig:comp_ex}}
\end{figure}

\begin{example}
Let us consider the model described in Example~\ref{ex:aebs}
and let us assume that the concrete input space $U$ of the CPSML model $M$ consists of the
initial velocity of the subject vehicle $vel(0)$, the initial
distance between the vehicle and the preceding obstacle $dist(0)$,
and the sequence of pictures that can be captured by the camera.
Let $\varphi := \G{[0,T]}(dist(t) \geq \tau)$ be a specification that 
requires the vehicle to be always 
farther than $\tau$ from the preceding obstacle.
Instead of analyzing the whole input space $U$ (including a vast number of pictures),
we adopt our compositional framework to target a 
specific subset of $U$. 
Let $M^+$ be the optimistic abstraction of the AEBS model, i.e., assuming
a perfectly working image classifier, 
and let $ml$ be the actual classifier. We begin by computing the
validity subsets $U_{\varphi}^+$ and $U_{\neg\varphi}^+$ of $\varphi$ against $M^+$, considering only $vel(0)$
and $dist(0)$ and assuming exact distance measurements during the
simulation. Next, we analyze only the image classifier $ml$
on pictures of obstacles whose distances fall in $U_{\varphi}^+$, say in 
a range $[d_m,d_M]$ (see Figure~\ref{fig:comp_ex}).
Our ML analyzer generates only pictures of obstacles whose distances are in
$[d_m,d_M]$, finds possible sets of images that are misclassified, and
returns the corresponding distances that, when projected back to $U$,
yields the subset $U^{ml}$ of $U$.
Finally, a temporal logic falsifier can be invoked to search over the restricted input space
$U^{ml}$ and a set of counterexamples is returned.

\end{example}

Algorithm~\ref{algo:CPSMLfalsification} and the above example 
show how our compositional approach relies on three key
steps:
(i) computing the validity domain for an STL formula for
a given simulation model;
(ii) falsifying an STL formula on a simulation model,
and 
(iii) a ML analyzer that computes a sub-space of its input
feature space that leads to misclassifications.
The first two steps have been well-studied in the literature
on simulation-based verification of CPS, and implemented in
tools such as Breach~\cite{donze2010breach}. 
We discuss our approach to Step (iii) in the next section
--- our ML analyzer
that identifies misclassifications of the ML component
relevant to the overall CPSML input space.


\section{Machine Learning Analyzer}
\label{sec:MLanalysis}

A central idea in our approach to analyzing CPSML models
is to use {\em abstractions} of the ML components. 
For instance, in the preceding section, we used the notions of
{\em perfect} ML classifiers and {\em always-wrong} classifiers
in computing the region of uncertainty (ROU).
In this section, we extend this abstraction-based approach to
the ML classifier and its input (feature) space.

One motivation for our approach comes from the application
domain of autonomous driving where machine learning is used for
object detection and perception. 
Instead of exploring the high-dimensional input space for
the ML classifier
involving all combinations of pixels, we instead perform the key
simplification of 
{\it exploring realistic and meaningful modifications} to a 
given image dataset that corresponds to the ROU.
Autonomous driving groups spend copious amounts of time collecting
images and video to train their learning-based perception systems with.
We focus on analyzing the space of images that is ``close'' to this
data set but with semantically significant modifications that can
identify problematic cases for the overall system.

The space of modifications to input feature vectors 
(say, images) induces an abstract space 
over the concrete feature (image) space.
Let us denote the abstract input domain by $\abssp$.
Given a classifier 
$\class : \featsp \to \labsp$, our ML analyzer computes a simpler 
function $\absclass : \abssp \to \labsp$ that approximates $\class$ on 
the abstract domain $\abssp$.
The abstract domain of the function $\absclass$ is analyzed and clusters of misclassifying
abstract elements are identified. The concretizations of such elements are subsets of features that are misclassified by the original classifier $\class$.
We describe further details of this approach in the remainder of this section.

\subsection{Feature Space Abstraction}
\label{sec:input_abs}


Let $\cfeatsp \subseteq \featsp$ be a subset of the feature space of
$\class : \featsp \to \labsp$.
Let $\leq$ be a total order on a set $\abssp$ called the abstract set. An abstraction function 
is an injective function $\absf : \cfeatsp \to \abssp$ that maps every feature vector $\feat \in \cfeatsp$
to an abstract element $\absf(\feat) \in \abssp$. Conversely, the concretization function
$\conf : \abssp \to \cfeatsp$ maps every abstraction $\va \in \abssp$ to a feature $\conf(\va) \in \cfeatsp$.

The abstraction and concretization functions play a fundamental
role in our falsification framework.
First, they allow us to map the input space of the CPS model to the
feature space of its classifiers. Second, the abstract space
can be used to analyze the classifiers on a compact domain
as opposite to intractable feature spaces.
These concepts are clarified in the following example, 
where a feature space of pictures is abstracted into a 
three-dimensional unit hyper-box.

\begin{example}\label{ex:abs}

Let $\featsp$ be the set of RGB pictures of size $1000\times 600$, i.e., $\featsp = \{0,\dots,255\}^{1000\times600\times3}$.
Suppose we are interested in analyzing a ML image classifier in the 
context of our AEBS system. In this case, we are interested in
images of road scenarios rather than on arbitrary images in $\featsp$. 
Further, assume that we start with a reference data set of images
of a car on a two-lane highway with a desert road background, as
shown in Figure~\ref{fig:X_and_A}.
Suppose that we are interested only in the constrained feature space
$\cfeatsp \subseteq \featsp$ comprising this desert road scenario with
a single car on the highway and three dimensions along which the
scene can be varied:
(i) the x-dimension (lateral) position of the car; 
(ii) the z-dimension (distance from the sensor) position
of the car, and (iii) the brightness of the image.
%
The $x$ and $z$ positions of the car
and the brightness level of the picture can be seen as the dimensions of 
an abstract set $\abssp$. In this setting, we can define the abstraction and concretization functions $\absf$ and $\conf$
that relate the abstract set $\abssp = [0,1]^3$ and $\cfeatsp$. For instance, the
picture $\conf(0,0,0)$ sees the car on the left, close to the observer, and low brightness;
the picture $\conf(1,0,0)$ places the car shifted to the right;
on the other extreme, $\conf(1,1,1)$ has the car on the right, far away from the observer, and  
with a high brightness level. Figure~\ref{fig:X_and_A} depicts some car pictures of $\tilde{\stsp}$
disposed accordingly to their position in the abstract domain $\abssp$ (the surrounding box).
	
	\begin{figure}
		\centering
		\includegraphics[scale=0.159]{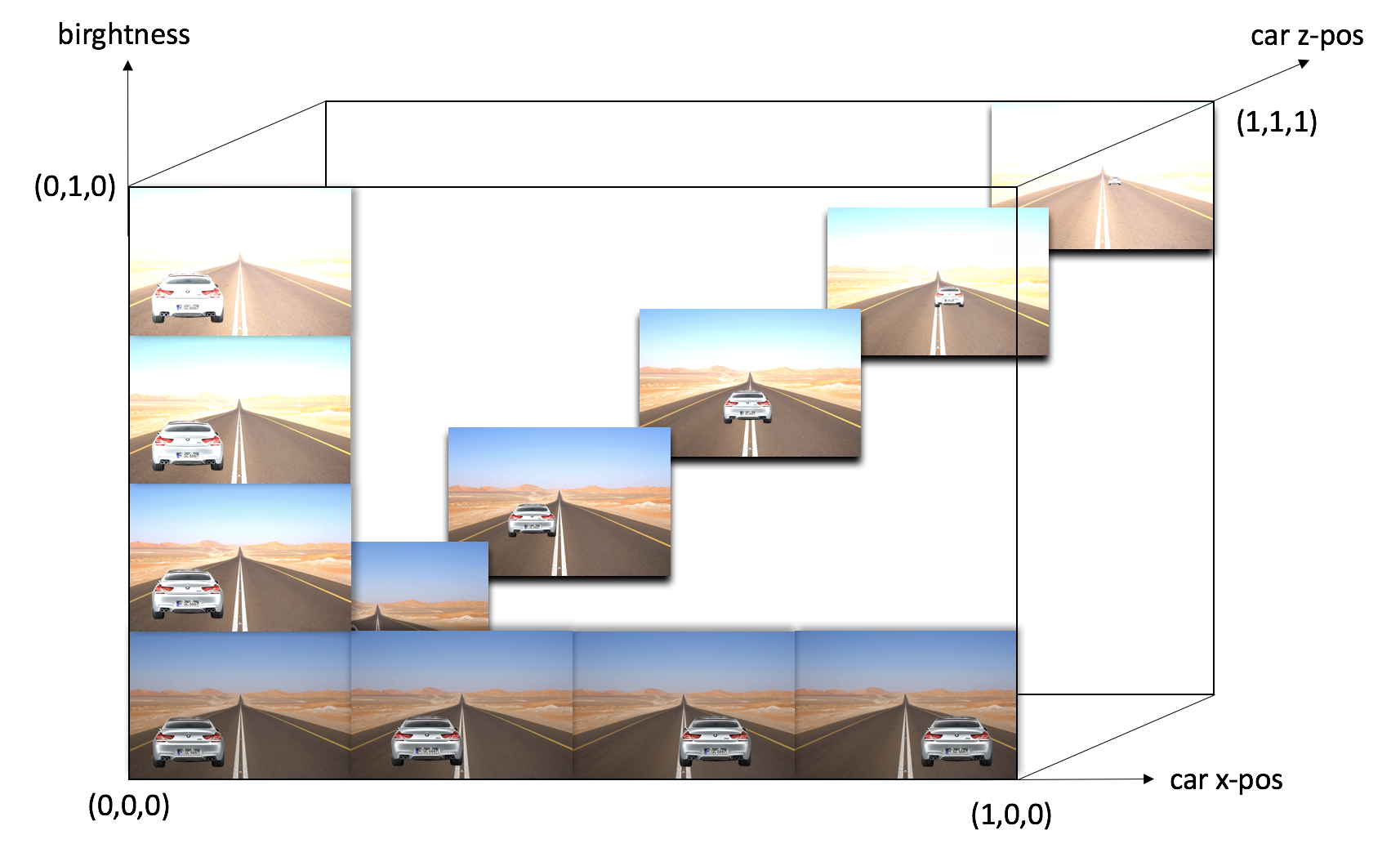}
		\caption{Feature Space Abstraction. The cube represents the abstract space $\abssp$ with the three dimensions corresponding to three different image modifications. The displayed road images correspond to concretized elements of the concrete feature space $\cfeatsp$.\label{fig:X_and_A}}
	\end{figure}
\end{example}

\subsection{Approximation of Learning Components}
\label{sec:approx_ML}

We now describe how the feature space abstraction can be used to construct an
approximation that helps the identification of misclassified feature vectors.

Given a classifier $\class : \featsp \to \labsp$ and a constrained feature space $\cfeatsp \subseteq \featsp$,
we want to determine an approximated classifier $\absclass : \abssp \to \labsp$, such that
$\err{\absclass}{T} \leq \epsilon$, for some $0 \leq \epsilon \leq 1$
and test set $T = \{ (\iafeat{1},\ilab{1}), \dots,  (\iafeat{l},\ilab{l}) \}$, with $\ilab{i} = \class(\conf(\iafeat{i}))$,
for $i = 1,\dots, l$.

Intuitively, the proposed approximation scheme samples elements from 
the abstract set, computes the labels of the concretized elements using the 
analyzed learning algorithm, and finally, interpolates the abstract elements
and the corresponding labels in order to obtain an approximation function.
Constructing such an approximation has multiple uses:
(i) The process of finding a good approximation of $\class$ appears to drive the sampling of elements
towards misclassifications that lead to system-level counterexamples,
and
(ii) the obtained approximation $\absclass$ can be used to perform higher-level analysis
of the reason for misclassifications, e.g., by identifying
clusters of misclassified feature vectors that share common characteristics.
We elaborate on these aspects later in this section.

\begin{algorithm}
	\caption{Approximation construction of classifier $\class : \featsp \to \labsp$}
    	\label{algo:approximation}
    	\begin{algorithmic}[1]
      		\Function{Approximation}{$\abssp,\conf, \epsilon$}\Comment{$\abssp$ abstract set ($\conf : \abssp \to \cfeatsp$), $0 \leq \epsilon \leq 1$ }
			\State $T_I \gets \emptyset$
			\Repeat
	 	 		\State $T_I \gets T_I \cup$ \Call{sample}{$\abssp,\class$}\label{ln:ti_samp}
				\State $\absclass \gets$ \Call{interpolate}{$T_I$}\label{ln:interp}
				\State $T_E \gets$ \Call{sample}{$\abssp,\class$}
			\Until{$\err{\absclass}{T_E} \leq \epsilon$}\label{ln:halt_con}
			\State \Return $\absclass$
      		\EndFunction
  	\end{algorithmic} 
\end{algorithm}

The \Call{Approximation}{} algorithm (Algorithm~\ref{algo:approximation}) formalizes the proposed approximation
construction technique. 
It receives in input an abstract domain $\abssp$ for the concretization
function $\conf : \abssp \to \cfeatsp$, with $\cfeatsp \subseteq \featsp$, the error
threshold $0 \leq \epsilon \leq 1$, and returns a function
$\absclass : \abssp \to \labsp$ that  approximates $\class$
on the constrained feature space $\cfeatsp$. The algorithm consists of a loop that
iteratively improves the approximation $\absclass$.
At every iteration, the algorithm populates the interpolation test set $T_I$
by sampling abstract features from $\abssp$ and computing the concretized 
labels according to $\class$ (Line~\ref{ln:ti_samp}), i.e.,
\Call{sample}{$\abssp,\class$}$ = \{ (\afeat,\lab) \mid \afeat \in \cabssp, \lab = \class(\conf(\afeat))\}$,
where $\cabssp \subseteq \abssp$ is a finite subset of samples determined with some 
sampling method.
Next, the algorithm interpolates the points of $T_I$ (Line~\ref{ln:interp}) according
to a suitably-chosen interpolation technique.
The result is a function $\absclass : \abssp \to \labsp$ that simplifies the
original classifier $\class$ on the concretized constrained feature space $\cfeatsp$.
The approximation is evaluated on the test set $T_E$.
Note that at each iteration, $T_E$ changes while $T_I$
incrementally grows. The algorithm iterates until the error rate $\err{\absclass}{T_E}$
is smaller than the desired threshold $\epsilon$ (Line~\ref{ln:halt_con}).
\revised{
This is a heuristic function approximation method, and there is no formal guarantee
on termination of the loop; however, in practice, we have found that the loop always
terminates with the desired error less than $\epsilon$.
}

The technique with which the samples in $T_E$ and $T_I$ are 
selected strongly influences the accuracy of the approximation.
\revised{A good sampling method can find corner-case misclassifications
and provide better high-level insight into the regions of the
abstract space where $\class$ generates misclassifications.}
In order to have a good coverage of the abstract set $\abssp$,
we propose the usage of low-discrepancy sampling methods that,
differently from uniform random sampling, cover sets quickly and evenly.
In this work, we use the Halton and lattice sequences, 
two common and easy-to-implement sampling methods, which we
explain next.

\subsection{Sampling Methods}
\label{sec:sample_methods}

Discrepancy is a notion from equidistribution theory~\cite{weyl1916gleichverteilung,rosenblatt1995pointwise} that finds application in 
quasi-Monte Carlo techniques for error estimation and approximating
the mean, standard deviation, integral,
global maxima and minima of complicated functions, 
such as, e.g., our classification functions.

\begin{definition}[Discrepancy~\cite{morokoff1994quasi}]
	Let $\featsp = \{  \ifeat{1}, \dots, \ifeat{m} \}$ be a finite set of points in
	$n$-dimensional unit space, i.e., $\featsp \subset [0,1]^n$. The \emph{discrepancy}
	of $\featsp$ is given by:
	\begin{equation}
		\disc{\featsp} = \sup_{B \in J} \mid \frac{\#(\featsp,B)}{m} - vol(B) \mid
	\end{equation}
	where $\#(\featsp,B) = |\{ \feat \in \featsp \mid \feat \in B \}|$, i.e., the number of
	points in $\featsp$ that fall in $B$, $vol(B)$ is the $n$-dimensional volume of $B$,
	and $J$ is the set of boxes of the form $\{ \feat \in \reals^n | a_i \leq \feat_i \leq b_i \}$,
	where $i=1,\dots,n$ and $0 \leq a_i < b_i < 1$.
\end{definition}

\begin{definition}[Low-discrepancy sequence~\cite{morokoff1994quasi}]
	A \emph{low-discrepancy sequence}, also called 
	\emph{quasi-random sequence}, is a sequence with the 
	property that for all $m \in \naturals$,
	its subsequence $\featsp = \{ \ifeat{1}, \dots, \ifeat{m} \}$ has low discrepancy.
\end{definition}

Low-discrepancy sequences fill spaces more uniformly than uncorrelated
random points.
This property makes low-discrepancy sequences suitable for problems where grids are involved, but it is unknown in advance how fine the grid must be to attain precise results.
A low-discrepancy sequence can be stopped 
at any point where convergence is observed, whereas the usual uniform random sampling technique requires a large number of computations between stopping points~\cite{trandafir_quasirandom}. 
Low-discrepancy sampling methods have
improved computational techniques in many areas, including
robotics~\cite{branicky2001quasi}, image processing~\cite{hannaford1993resolution}, computer graphics~\cite{shirley1991discrepancy}, numerical integration~\cite{sloan1994lattice}, and optimization~\cite{niederreiter1992random}.

We now introduce two low-discrepancy sequences that will
be used in this work. For more sequences and details see, e.g.,~\cite{niederreiter1988low}.

\begin{enumerate}
	\item \emph{Halton sequence}~\cite{morokoff1994quasi}.
Based on the choice of an arbitrary prime number $p$,
the $i$-th sample is obtained by representing $i$ in base $p$, reversing its digits, and 
moving the decimal point by one position. The resulting number is the $i$-th sample 
in base $p$. For the multi-dimensional case, it is sufficient to choose a different prime number for each dimension.
In practice, this procedure corresponds to choosing a prime base $p$, dividing the $[0,1]$ interval in $p$ segments,
then $p^2$ segments, and so on.
	\item \emph{Lattice sequence}~\cite{matousek2009geometric}. A lattice can be seen as the generalization of a multi-dimensional grid
with possibly nonorthogonal axes. Let $\alpha_1, \dots, \alpha_{n-1} \in \reals_{> 0}$ be irrational numbers
and $m \in \naturals$. The $i$-th sample of a lattice sequence is $(i/m,\{i\alpha_1\}, \dots,\{i\alpha_{n-1}\})$, where the curly braces  
$\{ \cdot \}$ denote the fractional part of the real value (modulo-one arithmetic). 
\end{enumerate}

\begin{example}\label{ex:MLan}

	We now analyze two Convolutional Neural Networks (CNNs):
	the Caffe~\cite{jia2014caffe} version of AlexNet~\cite{krizhevsky2012imagenet} and 
	the Inception-v3 model of Tensorflow~\cite{tensorflow2015},
	both trained on the ImageNet database~\cite{imagenet}.
	We sample $1000$ points from the abstract domain defined in Example~\ref{ex:abs}
	using the lattice sampling techniques.
	These points encode the $x$ and $z$ displacements
	of a car in a picture and its brightness level (see Figure~\ref{fig:X_and_A}).
	Figure~\ref{fig:mlan} (a) depicts the sampled points with their 
	concretized labels. The green circles indicate correct classifications,
	i.e., the classifier identified a car, the red circles denote misclassifications, i.e., no car detected.
	The linear interpolation of the obtained points
	leads to an approximation function. The error rates $\err{\absclass}{T_E}$ of the obtained
	approximations (i.e., the discrepancies between the predictions of the original image classifiers and their approximations)
	computed on $300$ randomly picked test cases are $0.0867$ and $0.1733$
	for AlexNet and Inception-v3, respectively.
	Figure~\ref{fig:mlan} (b) shows the projections of the approximation 
	functions for the brightness value $0.2$. The more red a region, 
	the larger the sets of pictures for which the neural networks do not 
	detect a car. For illustrative purposes, we superimpose the 
	projections of Figure~\ref{fig:mlan} (b) over the background used for the 
	picture generation. These illustrations show the regions of the concrete 
	feature vectors in which a vehicle is misclassified. 

\begin{figure}
\centering
	\subfloat[Sampling.\label{fig:sampling_lattice}]{
		\includegraphics[scale=0.25]{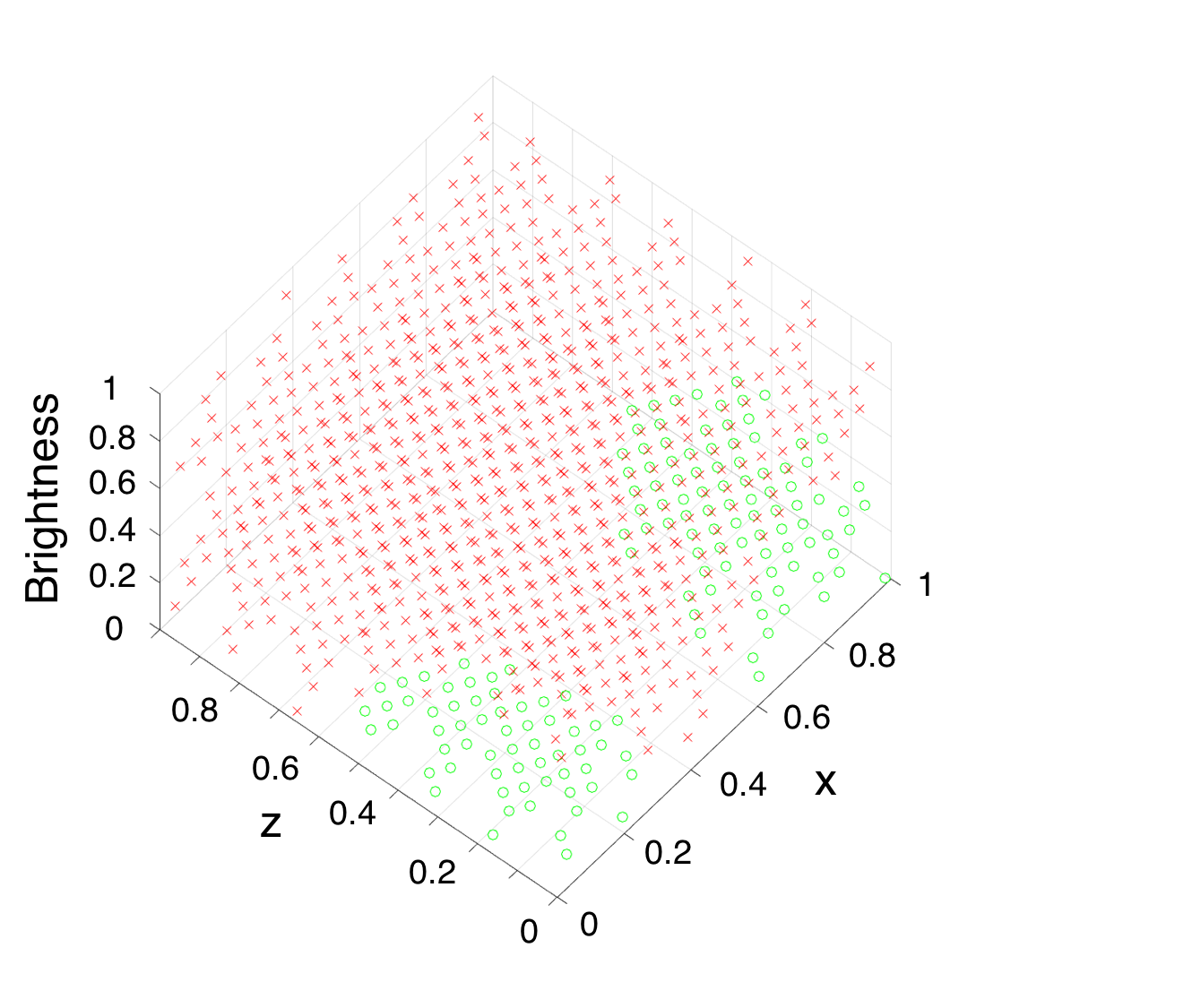}\qquad
		\includegraphics[scale=0.25]{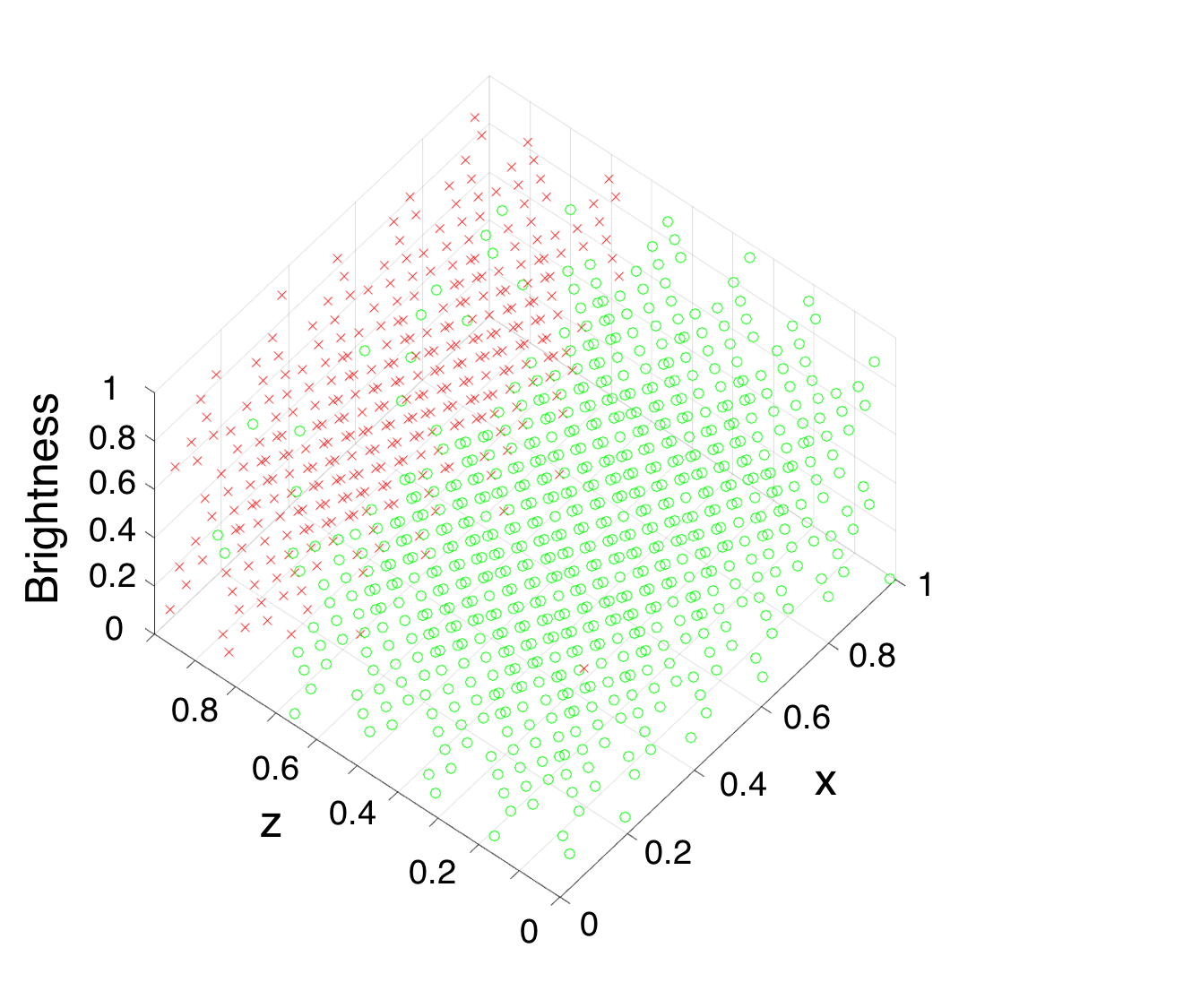}}\\
	\subfloat[Interpolation projection.\label{fig:interpolation_lattice}]{
		\includegraphics[scale=0.275]{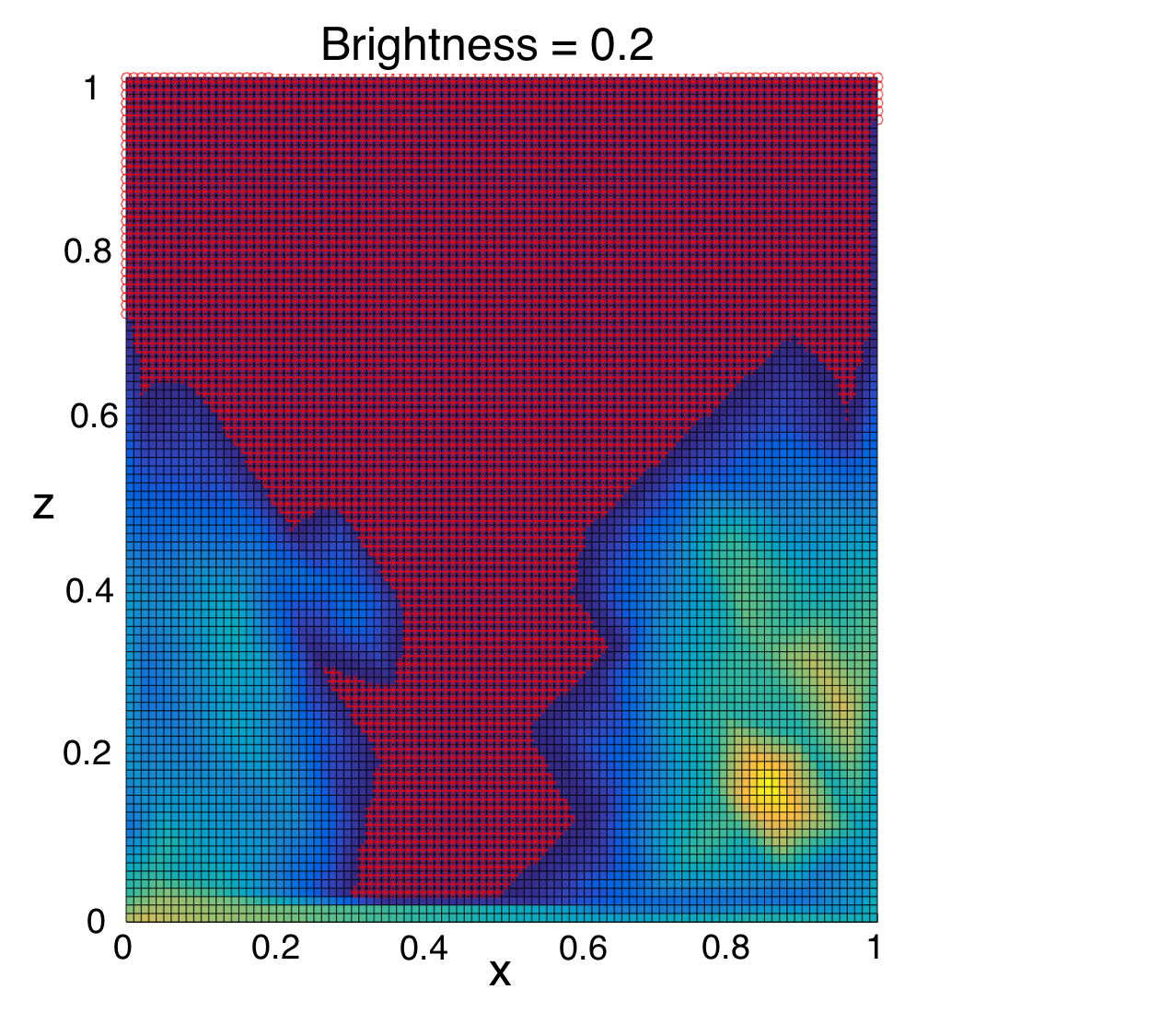}\qquad
		\includegraphics[scale=0.275]{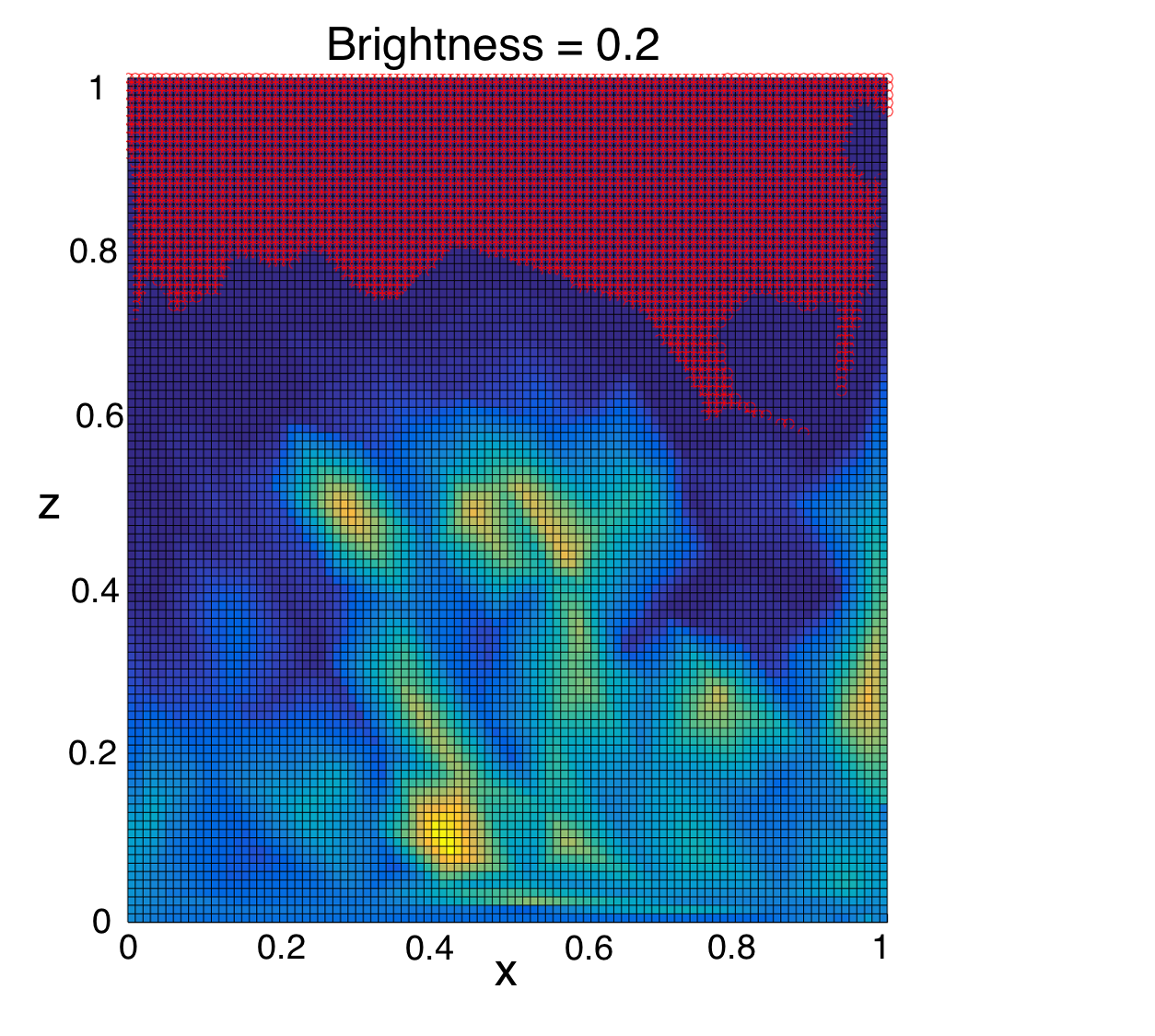}}\\
	\subfloat[Feature space analysis.\label{fig:analysis_inception_alexnet}]{
		\includegraphics[scale=0.4]{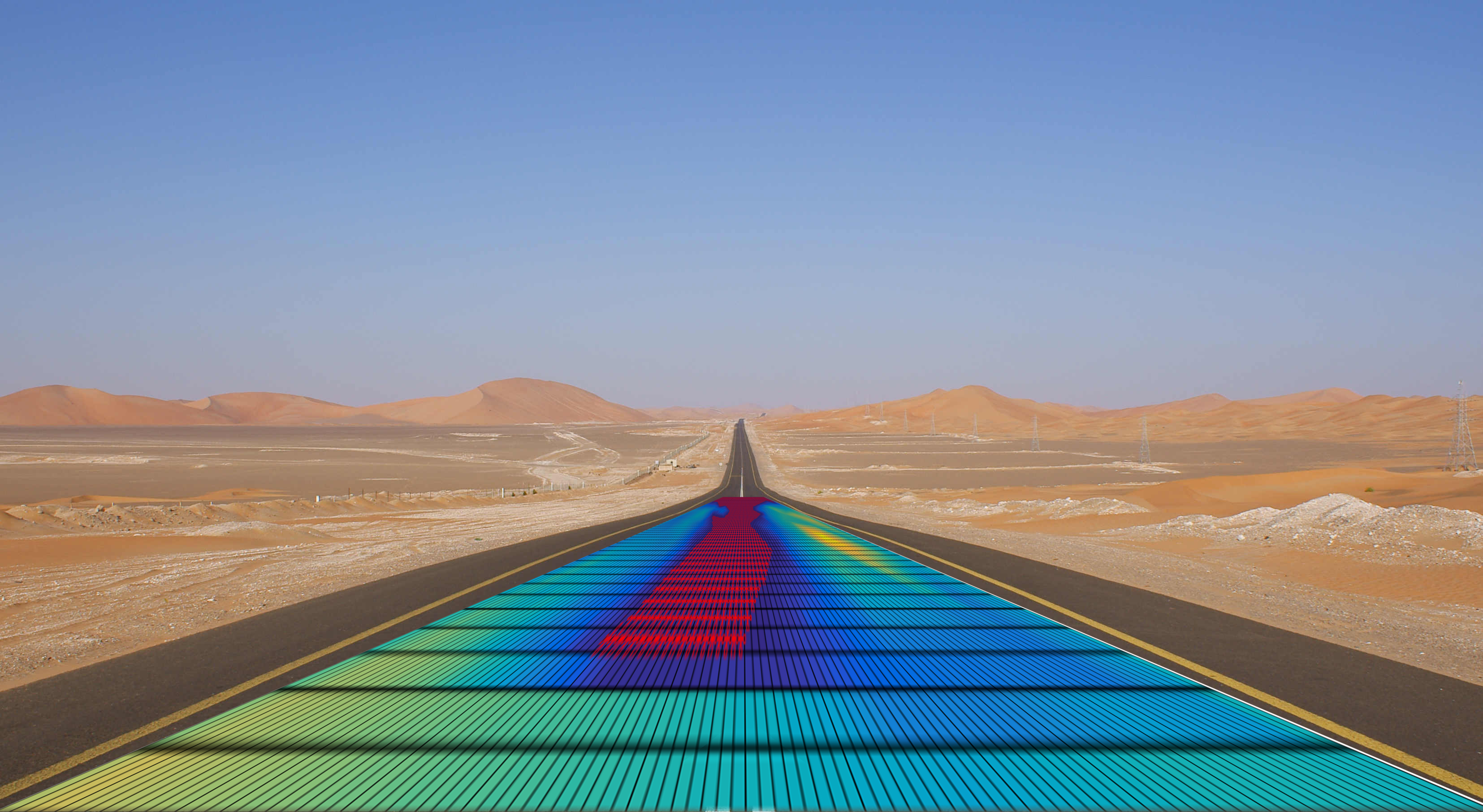}\qquad
		\includegraphics[scale=0.4]{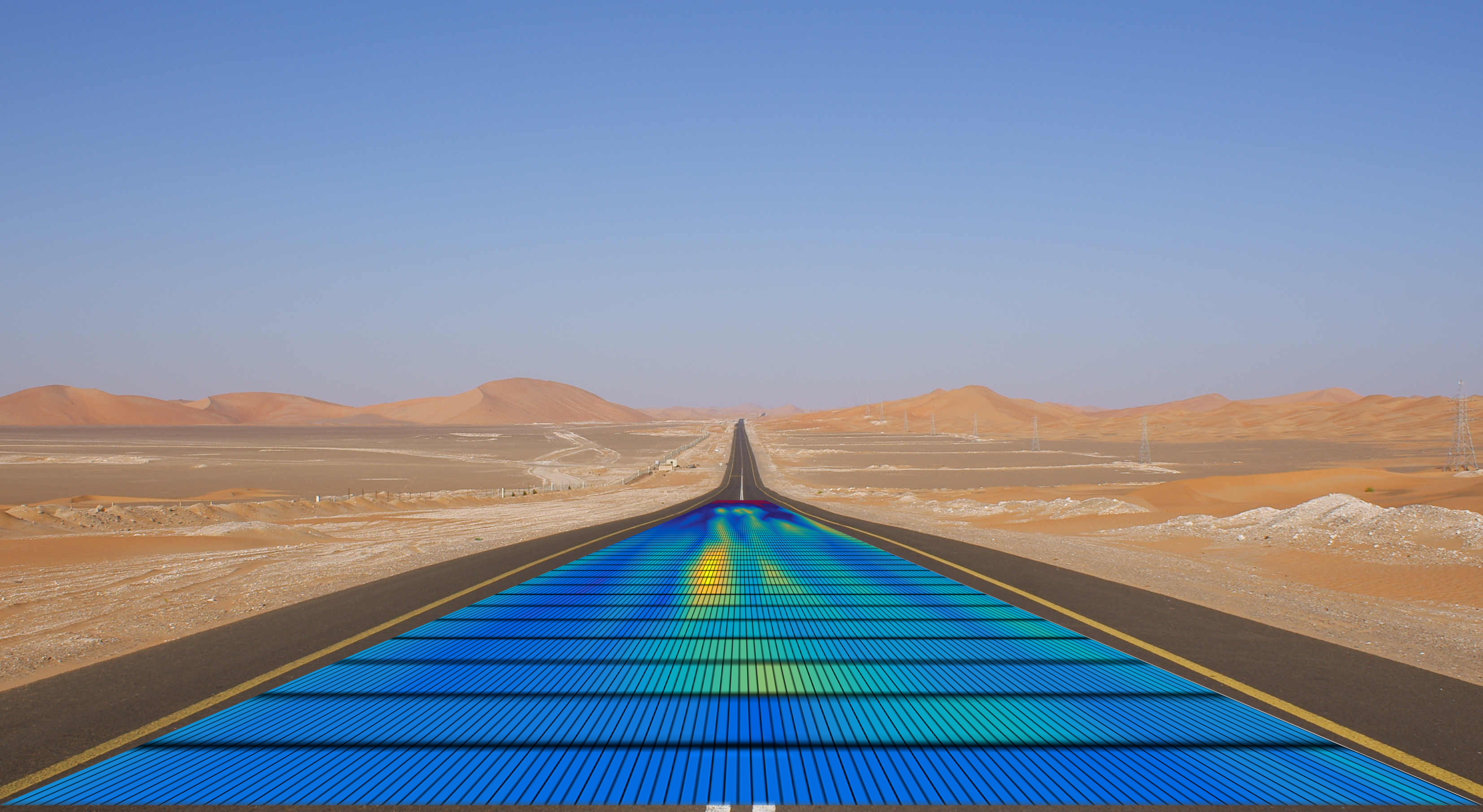}}
	\caption{ML analysis of AlexNet network developed with Caffe (top) and Inception-v3 network developed with Tensorflow (bottom) on a road scenario.\label{fig:mlan}}
\end{figure}

\end{example}

The analysis of Example~\ref{ex:MLan} on AlexNet and Inception-v3 provides useful insights.
First, we observe that Inception-v3 outperforms AlexNet on the considered road pictures since it correctly classifies
more pictures than AlexNet. Second, we notice that AlexNet tends to correctly classify pictures in 
which the $x$ abstract component is either close to $0$ or $1$, i.e., pictures in which the car is not in 
the middle of the street, but on one of the two lanes. This suggests that the model might not have been
trained enough with pictures of cars in the center of the road. Third, using the lattice method on Inception-v3,
we were able to identify a corner case misclassification in a cluster of correct predictions
(note the isolated red cross with coordinates $(0.1933 ,0.0244,0.4589)$).
All this information provides insights on the classifiers that can be useful in the hunt for counterexamples. 


\section{Experimental Results}
\label{sec:experiments}

\label{sec:exp:impl}

In this section we present two case studies, both involving
an Automatic Emergency Braking System (AEBS), but differing
in the details of the underlying simulator and controller. 
The first is a Simulink-based AEBS, 
the second is a Unity-Udacity simulator-based AEBS. 
The first case study showcases our full compositional falsification
method based on abstraction and compositional reasoning.
The second case study shows how we can apply falsification directly on
the concrete CPSML model (i.e. without performing the optimistic/pessimistic
abstractions of the neural network) while synthesizing
sequences of images that lead to a system-level counterexample.

The falsification framework for the first case study
has been implemented in a Matlab toolbox.\footnote{\url{https://github.com/dreossi/analyzeNN}}
The framework for the second case study has been written in Python 
and C\#.\footnote{\url{https://bitbucket.org/sseshia/uufalsifier}}
Our tools deal with models of CPSML and STL specifications.
They mainly consist of a temporal logic falsifier and an ML analyzer that interact to falsify the given STL specification against the
decomposed models.
As an STL falsifier, we chose the existing tool Breach~\cite{donze2010breach}, while the
ML analyzer has been implemented from scratch.
{\revised{
The main reason to choose Breach is that our system-level specification
is in signal temporal logic, which is the requirement language underlying
this tool. Further, it has been developed in our group at UC Berkeley,
allowing for better integration with the other software components.
We leverage both the qualitative and the quantitative semantics
of Breach. The qualitative semantics is used to compute
the validity domains of specifications used in our case studies.
The quantitative semantics is used by Breach (and similar tools) to
translate the STL formula into a cost function to be minimized so
as to find a trace that drives its value below zero, i.e., a 
property violation.
}}
The ML analyzer implementation has two components:
the feature space abstractor and the ML approximation algorithm (see Section~\ref{sec:MLanalysis}).
The feature space abstractor implements a scene generator that concretizes the abstracted feature vectors.
The algorithm that computes an approximation of the analyzed ML component gives the user the option of 
selecting the sampling method and interpolation technique, as well as setting the desired error rate.
Our tools are interfaced with the deep learning frameworks 
Caffe~\cite{jia2014caffe} and Tensorflow~\cite{tensorflow2015}.
We ran our tests on a desktop computer Dell XPS 8900, Intel (R) Core(TM) i7-6700 CPU 3.40GHz, DIMM RAM 16 GB 2132 MHz,
GPUs NVIDIA GeForce GTX Titan X and Titan Xp, with Ubuntu 14.04.5 LTS and Matlab R2016b.

\subsection{Case Study 1: Simulink-based AEBS}
\label{sec:case:simulink}

Our first case study is a closed-loop Simulink model of a semi-autonomous vehicle
with an Advanced Emergency Braking System (AEBS)~\cite{taeyoung2011development}
connected to a deep neural network-based image classifier.
The model mainly consists of a four-speed automatic transmission controller
linked to an AEBS that automatically prevents collisions with preceding obstacles and alleviates the
harshness of a crash when a collision is likely to happen (see Figure~\ref{fig:sim_model}).
\begin{figure}
	\centering
	\includegraphics[scale=0.16]{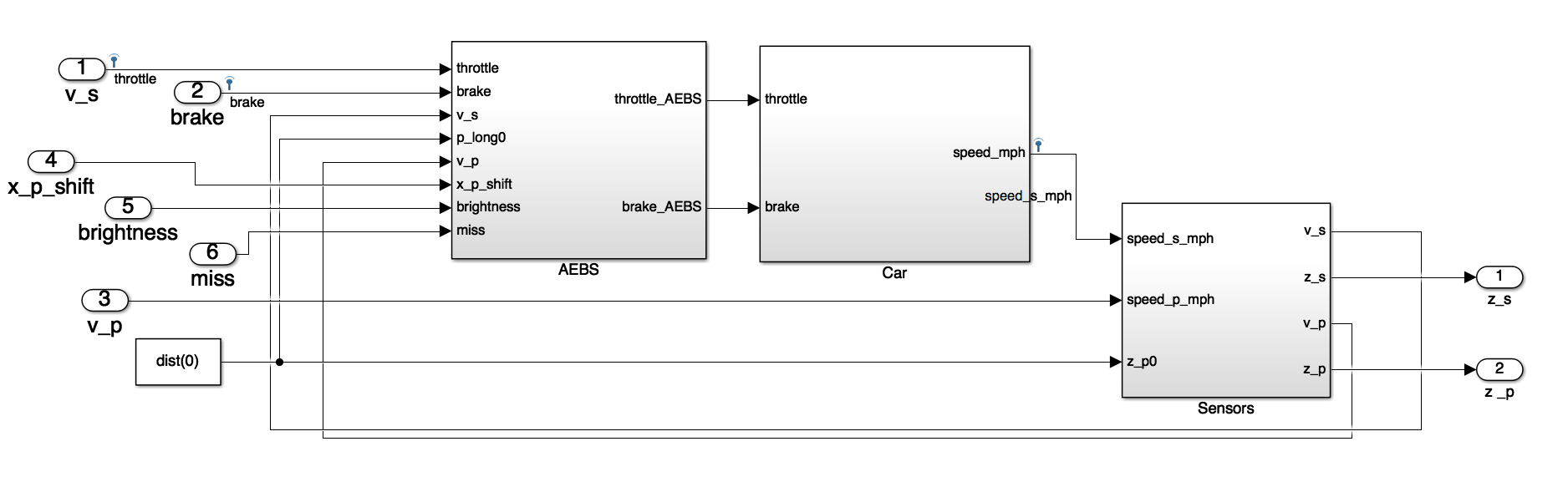}
	\caption{Simulink model of a semi-autonomous vehicle with AEBS. \label{fig:sim_model}}
\end{figure}
The AEBS determines a braking mode depending on the speed of the vehicle $v_s$,
the possible presence of a preceding obstacle, its velocity $v_p$, and the longitudinal distance $dist$ between the two.
The distance $dist$ is provided by radars having $30$m of range.
For obstacles farther than $30$m, the camera, connected
to an image classifier, alerts the AEBS that, in the case of detected obstacle, goes into warning mode.

Depending on $v_s, v_p, dist$, and the presence of obstacles detected by the image classifier,
the AEBS computes the time to collision and longitudinal safety indices, 
whose values determine a controlled transition between
safe, warning, braking, and collision mitigation modes. 
In safe mode, the car does not need to brake. 
In warning mode, the driver should brake to avoid a collision.
If this does not happen, the system goes into braking mode, where the 
automatic brake slows down the vehicle. 
Finally, in collision mitigation 
mode, the system, determining that a crash is unavoidable,
triggers a full braking action aimed to minimize the damage.

To establish the correctness of the system and in particular of its AEBS controller,
we formalize the STL specification $\G{}(\neg(dist(t)) \leq 0)$, that
requires $dist(t)$ to always be positive, i.e., no collision happens.
The input space is $v_s(0) \in [0,40]$ (mph), $dist(0) \in [0,60]$ (m), and the set of 
all RGB pictures of size $1000\times 600$. The preceding vehicle is not moving, i.e.,
$v_p(t) = 0$ (mph).

\begin{figure}
  \centering
  \includegraphics[width=\textwidth]{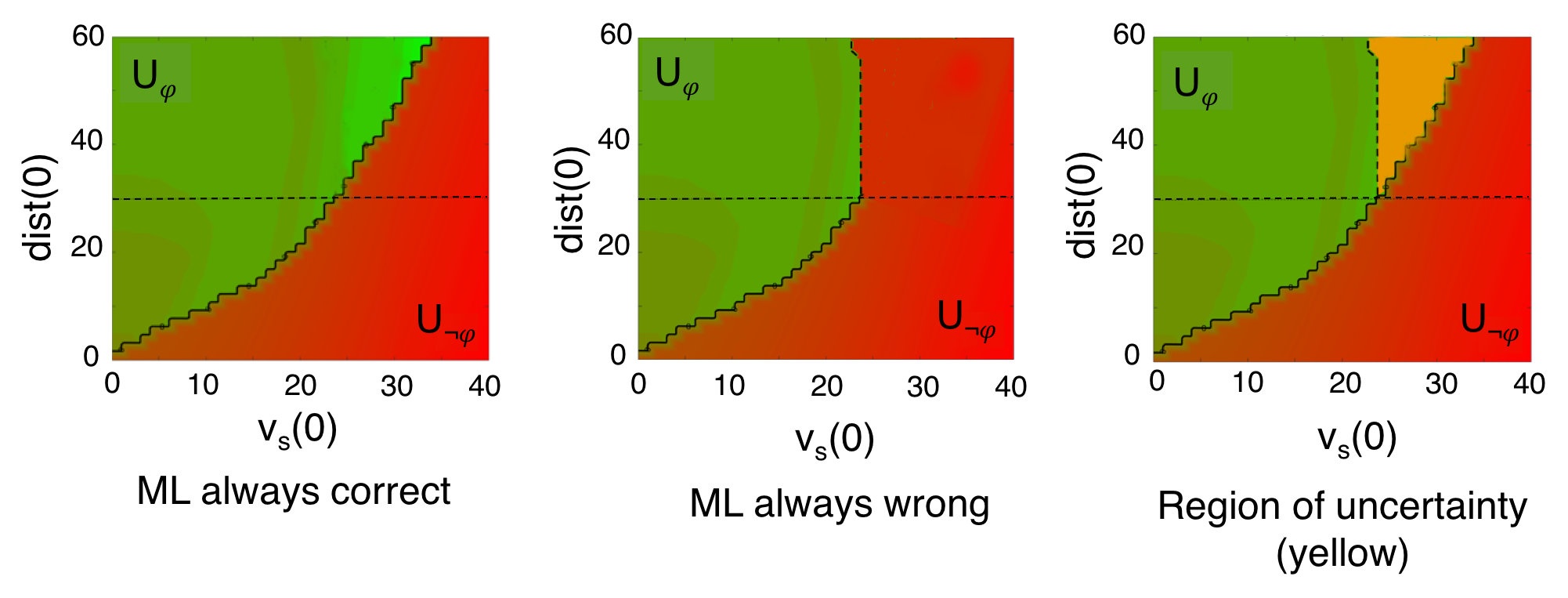}
	\caption{Validity domain for $\G{}(\neg(dist(t)) \leq 0)$ and AEBS model with different abstractions of ML component. {\footnotesize{The initial velocity and distance are on the x and y axes respectively.
The dotted (horizontal) line is the image classifier activation threshold.
Green indicates combinations of initial velocity and distance for which the property is satisfied and red indicates combinations for which the property is falsified. Our ML analyzer performs both optimistic (left) and pessimistic (middle) abstractions of the neural network classifier. 
	On the right-most image, the yellow region denotes the region of uncertainty (ROU).}}
\label{fig:ml_analysis_vis}}
\end{figure}

\begin{figure}
  	\centering
    	\includegraphics[width=0.5\textwidth]{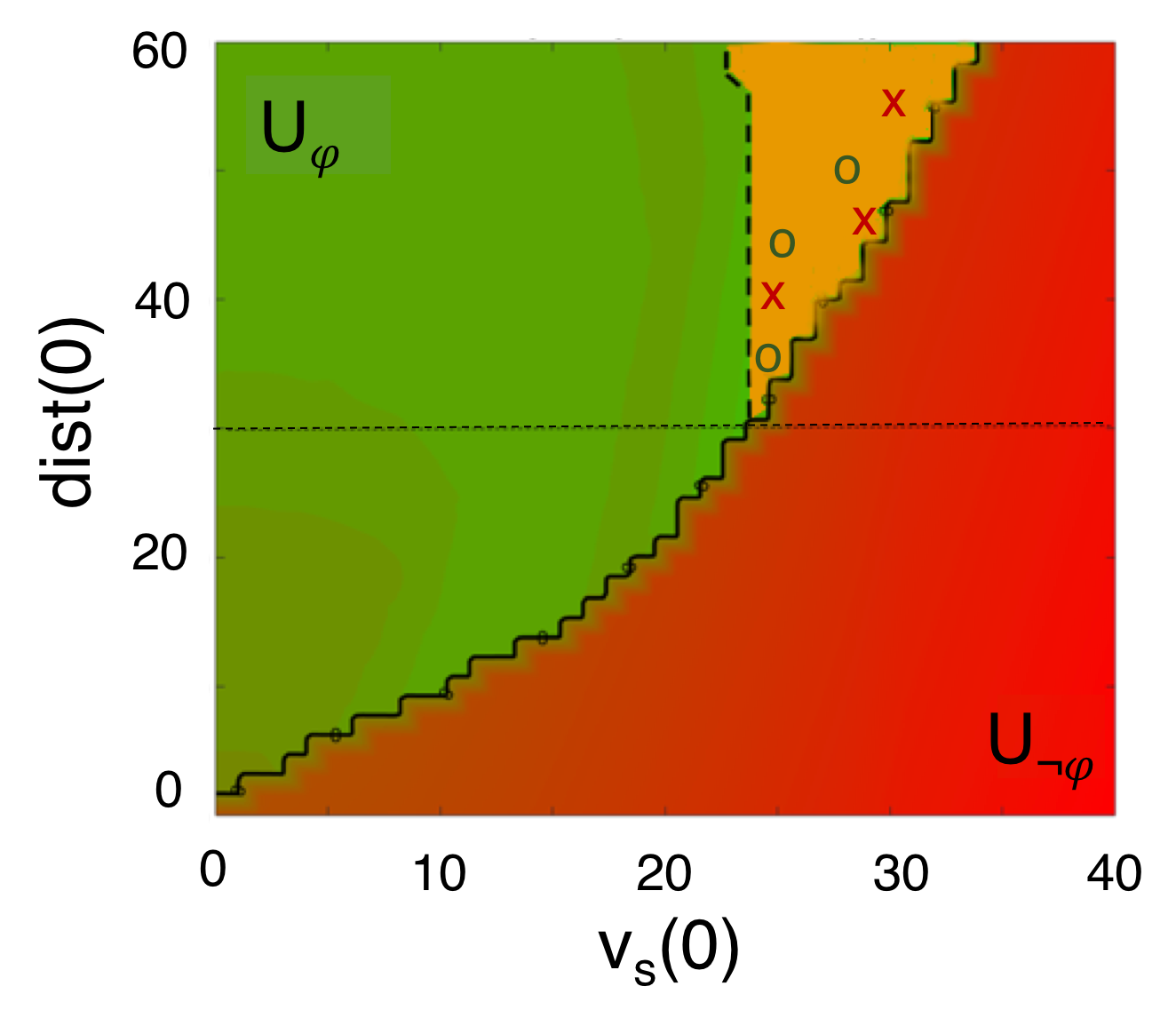}
	\caption{Analysis of Region of Uncertainty (ROU) for AEBS and property $\G{}(\neg(dist(t)) \leq 0)$. 
Red crosses in the ROU denote misclassifications generated by the ML analyzer that
leads to a system-level counterexample. A circle denotes a ``benign'' misclassification.
	\label{fig:no_miss}}
\end{figure}

At first, we compute the validity domain of $\varphi$ assuming that the radars are able to provide
exact measurements for any distance $dist(t)$ and the image classifier correctly detects the 
presence of a preceding vehicle. The computed validity domain 
is depicted in Figure~\ref{fig:ml_analysis_vis} (left-most image): green for $U_\varphi$
and red for $U_{\neg{\varphi}}$. Next, we try to identify candidate counterexamples 
that belong to the satisfactory set (i.e., the inputs that satisfy the specification) but might be influenced by a misclassification of the image classifier.
Since the AEBS relies on the classifier only for distances larger than $30$m, we can focus on the
subset of the input space with $dist(0) \geq 30$. Specifically, we identify potential counterexamples
by analyzing a pessimistic version of the model where the ML component always misclassifies 
the input pictures (see Figure~\ref{fig:ml_analysis_vis}, middle image).
From these results, we can compute the region of uncertainty, shown in
Figure~\ref{fig:ml_analysis_vis} on the right.
We can then focus our attention on the ROU, as shown in Fig.~\ref{fig:no_miss}.
In particular, we can identify candidate counterexamples,
such as, for instance, $(25,40)$ (i.e., $v_s(0) = 25$ and $dist(0) = 40$).


Next, let us consider the AlexNet image classifier and the ML analyzer presented in Section~\ref{sec:MLanalysis} that generates
pictures from the abstract feature space $\abssp = [0,1]^3$, where the dimensions of $\abssp$
determine the $x$ and $z$ displacements of a car and the brightness of a generated picture, respectively.
The goal now is to determine an abstract feature $\va_c \in \abssp$ related to the candidate counterexample $(25,40)$,
that generates a picture that is misclassified by the ML component and might lead to a violation of the specification $\varphi$.
The $dist(0)$ component of $\vu_c = (25,40)$ determines a precise $z$ displacement $\va_2 = 0.2$ in the abstract picture.
{\revised{The connection between the abstract and input spaces is defined by the abstraction function that, in this case,
was manually defined by the user. In general, this connection can be explicitly provided by a synthetic data generator such as, for instance, a simulator or image renderer.}
Now, we need to determine the values of the abstract $x$ displacement and brightness. Looking at the interpolation
projection of Figure~\ref{fig:mlan} (b), we notice that the approximation function misclassifies pictures with abstract component $\va_1 \in [0.4,0.5]$ and $\va_3 = 0.2$. Thus, it is reasonable to try to falsify
the original model on the input element $v_s(0) = 25, dist(0) = 40$, and concretized picture $\conf(0.5,0.2,0.2)$.
For this targeted input, the temporal logic falsifier computed a robustness value for $\varphi$ of $-24.60$, meaning that a falsifying counterexample
has been found. Other counterexamples found with the same technique are, e.g.,  
$(27,45)$ or $(31,56)$ that, associated with the correspondent concretized pictures with $\va_1 = 0.5$ and $\va_3 = 0.2$, lead to the robustness values  $-23.86$ and $-24.38$, respectively (see Figure~\ref{fig:no_miss}, red crosses). Conversely, we also disproved some candidate counterexamples,
such as $(28,50)$, $(24,35)$, or $(25,45)$, whose robustness values are $9.93, 7.40$, and $7.67$ (see Figure~\ref{fig:no_miss}, green circles).

For experimental purposes, we try to falsify a counterexample in which we change the $x$ position of the abstract feature so that
the approximation function correctly classifies the picture. For instance, by altering the counterexample $(27,45)$ with 
$\conf(0.5,0.225,0.2)$ to $(27,45)$ with $\conf(1.0,0.225,0.2)$, we obtain a robustness value of $9.09$, that means that the AEBS
is able to avoid the car for the same combination of velocity and distance of the counterexample, but different $x$ position of the preceding vehicle.
Another example, is the robustness value $-24.38$ of the falsifying input $(31,56)$ with $\conf(0.5,0.28,0.2)$, that altered to $\conf(0.0,0.28,0.2)$,
changes to $12.41$.

Finally, we test Inception-v3 on the corner case misclassification identified in Section~\ref{sec:approx_ML}
(i.e., the picture $\conf(0.1933,0.0244,0.4589)$). The distance $dist(0) = 4.88$ related to this abstract feature is below the activation
threshold of the image classifier. Thus, the falsification points are exactly the same as those of the computed validity domain (i.e., $dist(0) = 4.88$
and $v_s(0) \in [4,40]$). This study shows how a misclassification of the ML component might not affect the correctness of the CPSML model.

\subsection{Case Study 2: Unity-Udacity Simulator-based AEBS}
\label{sec:case:uubsim}

We now analyze an AEBS deployed within Udacity's self-driving car simulator.\footnote{Udacity's Self-Driving Car Simulator: \url{https://github.com/udacity/self-driving-car-sim}} The simulator,
built with the Unity game engine\footnote{Unity: \url{https://unity3d.com/}},  can be used to teach cars how to navigate roads using deep learning. We modified the simulator
in order to focus exclusively on the braking system. In our settings, the car steers by following some predefined waypoints, while 
acceleration and braking are controlled by an AEBS connected to a CNN.
An onboard camera sends images to the CNN whose
task is to detect cows on the road. Whenever an obstacle is detected, the AEBS triggers a brake that slows the vehicle down and prevents the collision against the obstacle. 

We implemented a CNN that classifies the pictures captured by the onboard camera in two categories
\lq\lq cow\rq\rq\ and \lq\lq not cow\rq\rq.
The CNN has been implemented and trained using Tensorflow.
We connected the CNN to the Unity C\# class that controls the car. The communication between the neural network and the
braking controller happens via Socket.IO protocol.\footnote{Socket.IO protocol: \url{https://github.com/socketio}}
A screenshot of the car braking in presence of a cow is shown in Figure~\ref{fig:cow_no_collision}.
A video of the AEBS in action can be seen at \url{https://www.youtube.com/watch?v=Sa4oLGcHAhY}.

\begin{figure}
	\resizebox{\textwidth}{!}{
	\subfloat[Correct detection and braking.\label{fig:cow_no_collision}]{\includegraphics[scale=0.3]{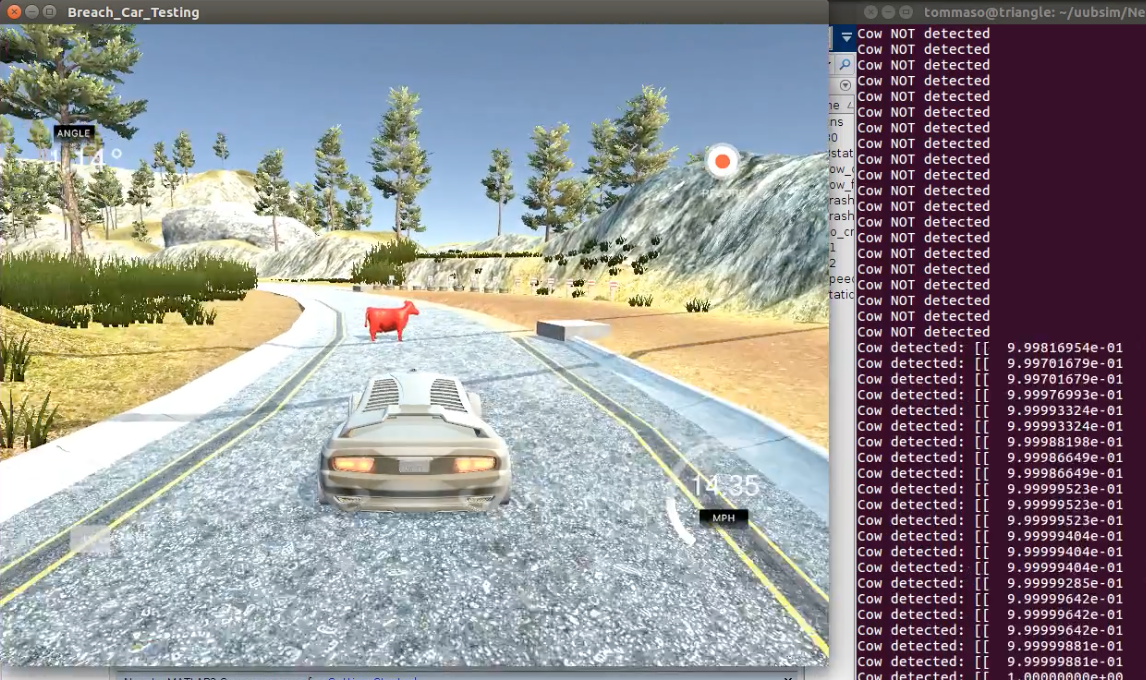}}
	~
	\subfloat[Misclassification and collision.\label{fig:cow_collision}]{\includegraphics[scale=0.1515]{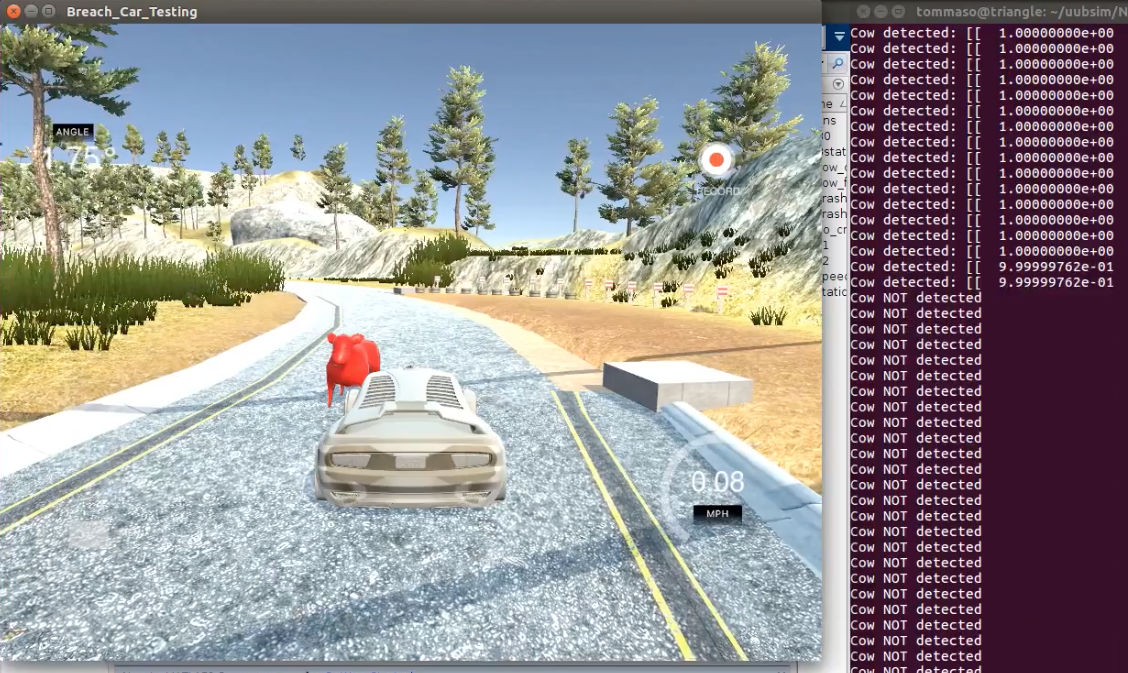}}
	}
	\caption{Unity-Udacity simulator AEBS. The onboard camera sends images to the CNN. When a cow is detected
	a braking action is triggered until the car comes to a complete stop.
	Full videos available at \protect\url{https://www.youtube.com/watch?v=Sa4oLGcHAhY} and \protect\url{https://www.youtube.com/watch?v=MaRoU5OgimE}. \label{fig:unity_aebs}}
\end{figure}

The CNN architecture is depicted in Figure~\ref{fig:nn}.
The network consists of eight layers: the first six are alternations of convolutions and max-pools with ReLU activations,
the last two are a fully connected layer and a softmax that outputs the network prediction.
The dimensions and hyperparameters of our neural network are
shown in Table~\ref{tab:nn_params}, where $l$ is a layer, $n_{H}^{[l]} \times n_{W}^{[l]} \times n_{C}^{[l]}$ is the dimension of the volume computed by the layer $l$,
$f^{[l]}$ is the filter size, $p^{[l]}$ is the padding, and $s^{[l]}$ is the stride.

\begin{figure}
	\centering
	\includegraphics[scale=0.25]{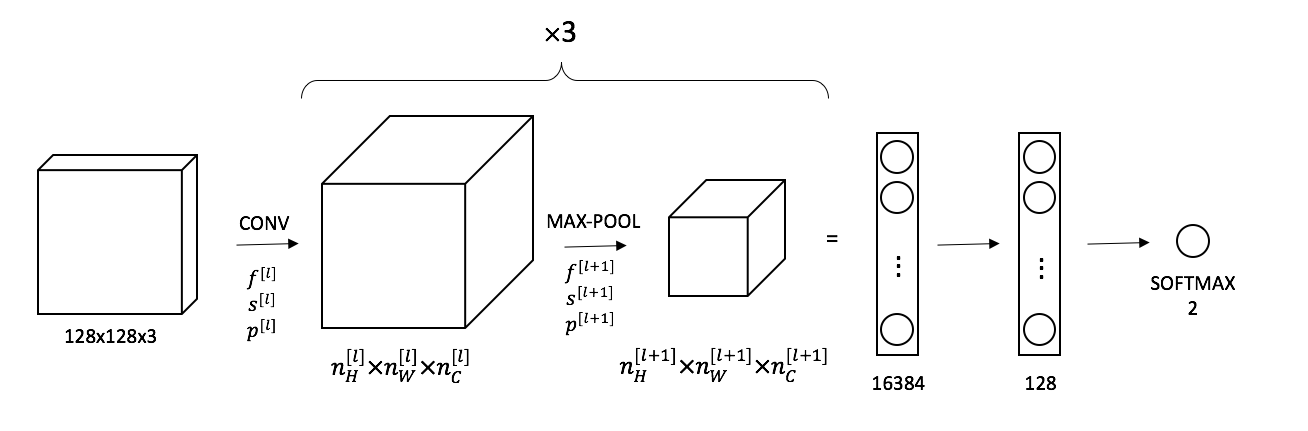}
	\caption{CNN architecture. \label{fig:nn}}
\end{figure}

\begin{table}
	\centering
	\resizebox{\textwidth}{!}{
	\begin{tabular}{| c | c c c c c c c c c |}
		\hline
		 & 0 & 1 & 2 & 3 & 4 & 5 & 6 & 7 & 8\\
		 \hline
		 $n_{H}^{[l]} \times n_{W}^{[l]} $	& $128 \times 128$ & $128 \times 128$ & $64 \times 64$ & $64 \times 64$ & $32 \times 32$ & $32 \times 32$ & $16 \times 16$ & $128 \times 1$ & $2 \times 1$	\\
		 $n_{C}^{[l]}$				& 3 & 32 & 32 & 32 & 32 & 64 & 64 & 1 & 1\\
		 $f^{[l]}$ & - & 3 & 2 & 3 & 2 & 3 & 2 & - & - \\
		 $p^{[l]}$ & - & 1 & 0 & 1 & 0 & 1 & 0 & - & - \\
		 $s^{[l]}$ & - & 1 & 2 & 1 & 2 & 1 & 2 & - & - \\
		 \hline
	\end{tabular}}
	\caption{CNN dimensions and hyperparameters.\label{tab:nn_params}}
\end{table}

Our dataset, composed by $1$k road images, was split into $80\%$ train data and $20\%$ validation.
We trained our model using cross-entropy cost function and Adam algorithm optimizer with learning rate $10^{-4}$. Our model 
reached $0.95$ accuracy on the validation set.

In our experimental evaluation, we are interested in finding a case where our AEBS fails, i.e., the car collides against a cow.
This requirement can be formalized as the STL specification 
$\G{}( \norm{\vx_{car}- \vx_{cow}} > 0 )$ that imposes the Euclidean distance of the car and cow positions
($\vx_{car}$ and $\vx_{cow}$, respectively) to be always positive. 

We analyzed the CNN feature space by considering the abstract space $\abssp = [0,1]^3$, where the dimensions of $\abssp$
determine the displacement of the cow of $\pm 4$m along the $x$-axis, its rotation along the $y$-axis, and the intensity of the red color channel.
We sampled the elements from the abstract space using both Halton sequence and a grid-based approach. The obtained results are shown
in Figure~\ref{fig:cnn_cow_analysis}. In both figures, green points are 
those that lead to images that are correctly classified by the CNN;
conversely, red points denote images that are misclassified by the CNN and can potentially lead to a system falsification. Note how we were able
to identify a cluster of misclassifying images (lower-left corners of both Figures~\ref{fig:grid_cow} and~\ref{fig:halton_cow}) as well as an isolated corner case (upper-center, Figure~\ref{fig:grid_cow}).

\begin{figure}
	\subfloat[Grid-based sampling.\label{fig:grid_cow}]{\includegraphics[scale=0.35]{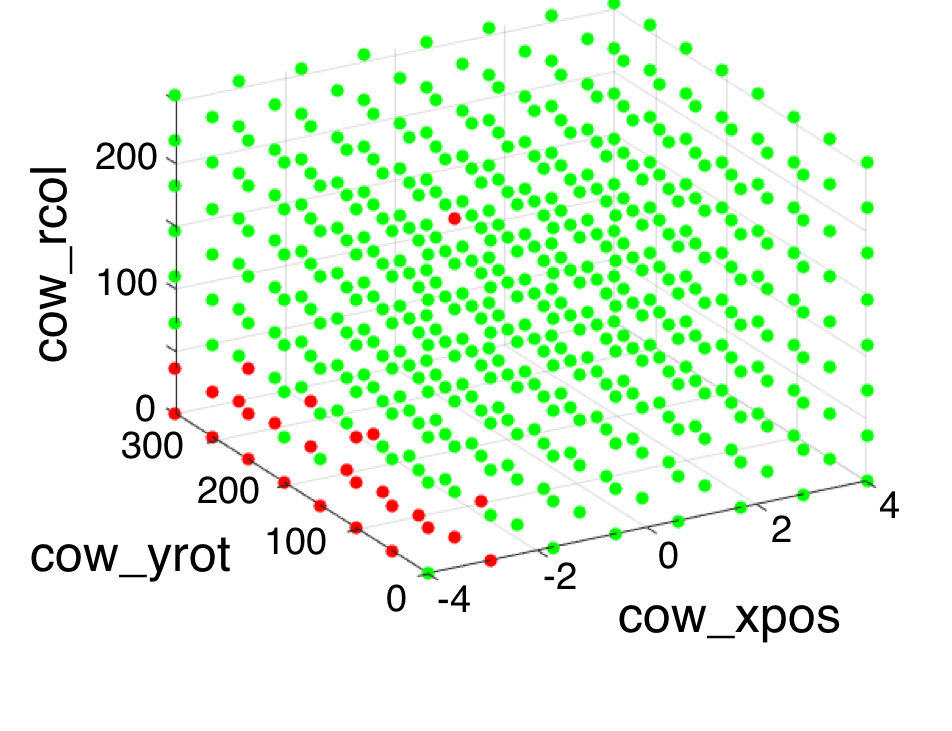}}
	\subfloat[Halton sequence sampling.\label{fig:halton_cow}]{\includegraphics[scale=0.35]{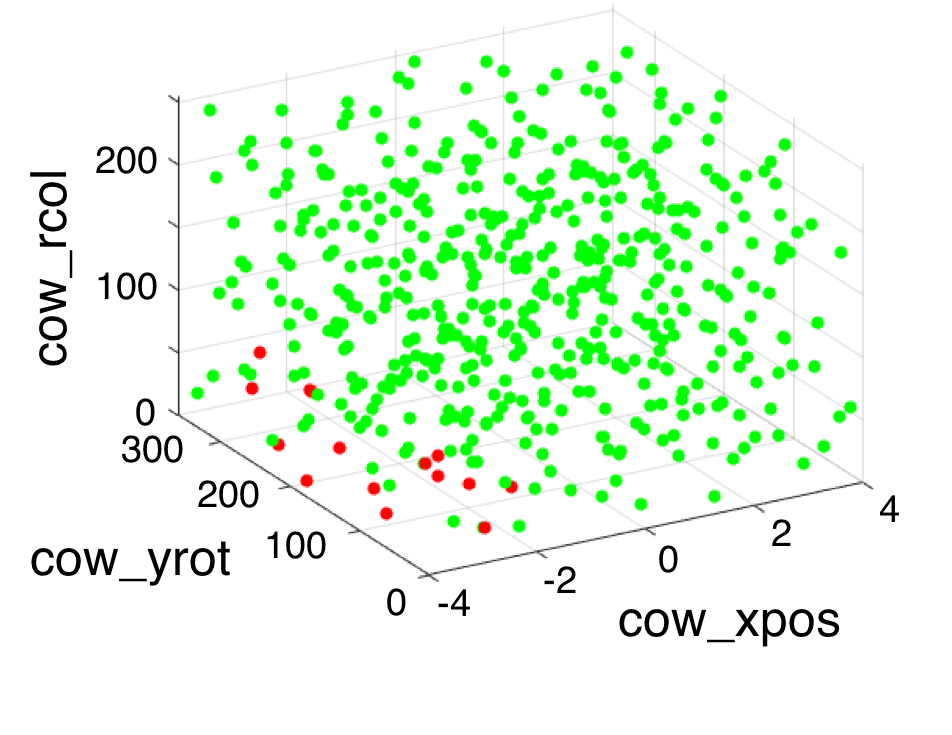}}
	\caption{CNN analysis.\label{fig:cnn_cow_analysis}}
\end{figure}

Finally, we ran some simulations with the misclassifying images 
identified by our analysis. Most of them brought the car to collide
against the cow. A screenshot of a collision is shown in Figure~\ref{fig:cow_collision}.
The full video is available at~\url{https://www.youtube.com/watch?v=MaRoU5OgimE}.

\section{Conclusion}\label{sec:conclusion}

We presented a compositional falsification framework for STL specifications against CPSML models
based on a decomposition between the analysis of machine learning
components and the system contained them.
We introduced an ML
analyzer able to abstract feature spaces, approximate ML classifiers, 
and provide sets of misclassified feature vectors that can be used to drive the falsification process.
We implemented our framework and showed its effectiveness for an
autonomous driving controller using perception based on deep neural networks.

This work lays the basis for future advancements. There are several
directions for future work, both theoretical and applied. In the
remainder of this section, we describe this landscape for future work.
See~\cite{SeshiaS16} for a broader discussion of these points in
the context of the goal of verified intelligent systems.

\noindent
{\em Improvements in the ML Analyzer:} 
We intend to improve our ML Analyzer
exploring the automatic generation of feature space abstractions from given
training sets. One direction is to exploit the structure of ML
components, e.g., the custom architectures that have been
developed for deep neural networks in applications such as autonomous
driving~\cite{iandola2016squeezenet}.
For instance, one could perform a sensitivity analysis that 
indicates along which axis in the abstract
space we should move in order to change the output label or reduce
the confidence of the classifier on its output.
Another direction is to improve the sampling techniques
that we have explored so far, ideally devising one that
captures the probability of detecting a corner-case scenario leading
to a property violation. Of particular interest are adaptive sampling methods
involving further cooperation between the ML Analyzer and the CPS Analyzer. 
We are also interested in integrating other techniques
for generating misclassifications of ML components 
(e.g.,~\cite{moosavi2015deepfool,huang-arxiv16,carlini-ieeesp17})
into our approach.

\noindent
{\em Impacting the ML component design:}
Our falsification approach produces input sequences that result
in the violation of a desired property.
While this is useful, it is arguably even more useful to obtain higher-level
interpretable insight into where the training data falls short, what
new scenarios must be added to the training set, and how the 
learning algorithms' parameters must be adjusted to improve accuracy.
For example, one could use techniques for mining specifications
or requirements (e.g.,~\cite{jin-tcad15,vazquez-cav17}) 
to aggregate interesting test images or video
into a cluster that can be represented in a high-level fashion.
One could also apply our ML Analyzer outside the 
falsification context, such as for controller synthesis. 

\noindent
{\em From Falsification to Verification:}
The compositional framework we introduced in this paper can also be used
for verification of a CPSML model. In particular, if one has a verifier
that can prove safety properties of the abstract systems (the optimistic
and pessimistic abstractions shown as $M^-$ and $M^+$ in Figure~\ref{fig:compFalsify})
and one has a systematic refinement strategy that can cover the
ROU, 
then the same compositional framework can be used to prove safety
properties of the CPSML model. We plan to explore this direction
in future work.

\noindent
{\em Further Applications:}
Although our approach has shown initial promise for reasoning about
autonomous driving systems, much more remains to be done to make this
practical. Real sensor systems for autonomous driving involve multiple sensors
(cameras, LIDAR, RADAR, etc.) whose raw outputs are often fused and 
combined with deep learning or other ML techniques to
extract higher level information (such as the location and type
of objects around the vehicle). This sensor space has very high
dimensionality and high complexity, not to mention 
streams of sensor input (e.g., video), that one must be able to analyze
efficiently.
To handle industrial-scale production systems, our overall analysis
must be scaled up substantially,
potentially via use of cloud computing infrastructure.
Finally, our compositional 
methodology could be extended to other, non-cyber-physical,
systems that contain ML components.

\bibliographystyle{abbrv}
\bibliography{refs}

\end{document}